\documentclass[journal,final]{IEEEtran}

\usepackage{cite}      
\usepackage{graphicx}  
\usepackage{overpic}    
\usepackage{subfigure} 
\usepackage{url}       
\usepackage{stfloats}  
\usepackage{amsmath}   
\usepackage{amssymb}   
\usepackage{epstopdf}

\newcommand{\D}{\displaystyle}
\usepackage{amsmath}
\usepackage{amsthm}
\usepackage{amssymb}
\usepackage{cite}
\usepackage{graphicx}
\usepackage{booktabs}
\usepackage{subfigure}

\newcommand{\q}{\mathsf{q}}

\newtheorem{theo}{Theorem}

\newtheorem{lemma}{Lemma}

\begin{document}

\title{Uplink Multicell Processing with Limited Backhaul via 
Per-Base-Station Successive Interference Cancellation 
\thanks{Manuscript submitted to
IEEE Journals on Selected Areas of Communications on Oct 13, 2012, revised on
March 20, 2013, accepted on May 2, 2013.  This
work was supported by the Natural Science and Engineering Research
Council (NSERC) of Canada. The materials in this paper have been presented in part
at IEEE Global Communications Conference (Globecom), Anaheim, CA, U.S.A., Dec 2012. }
\thanks{ L. Zhou was with the The Edward S. Rogers Sr. Department of
Electrical and Computer Engineering, University of Toronto, Toronto, ON M5S
3G4 Canada. He is now with Qualcomm Technologies Inc., Santa Clara, CA
95051 USA (email: zhoulei@comm.utoronto.ca).}
\thanks{
W. Yu is with  The Edward S. Rogers Sr. Department of Electrical
and Computer Engineering, University of Toronto, Toronto, ON M5S 3G4
Canada (email: weiyu@comm.utoronto.ca). Phone: 416-946-8665. FAX: 416-978-4425.
Kindly address correspondence to Wei Yu.}
}

\author{Lei Zhou, \IEEEmembership{Member, IEEE} 
	and Wei Yu, \IEEEmembership{Senior Member, IEEE}
}

\markboth{Submitted to IEEE Journals on Selected Areas in Communications}
{Uplink Multicell Processing with Limited Backhaul via Per-Base-Station 
Successive Interference Cancellation}

\maketitle

\thispagestyle{empty}

\begin{abstract}
This paper studies an uplink multicell joint processing model in which
the base-stations are connected to a centralized processing server via
rate-limited digital backhaul links. Unlike previous studies where the
centralized processor jointly decodes all the source messages from all
base-stations, this paper proposes a simple scheme which
performs Wyner-Ziv compress-and-forward relaying on a per-base-station 
basis followed by successive interference cancellation (SIC) at the
central processor. The proposed scheme significantly reduces the
implementation complexity of joint decoding while resulting in an
easily computable achievable rate region. 
Although suboptimal in general, this paper proves that the proposed
per-base-station processing SIC scheme can achieve the sum capacity 
of a special cellular Wyner model to within a constant gap.
Under the proposed SIC framework, this paper also studies the capacity
scaling with limited backhaul. It is established that in order to
achieve to within constant gap to the maximum SIC rate with
infinite backhaul, the limited-backhaul system must have 
backhaul link capacities that scale logarithmically with the 
signal-to-interference-and-noise ratios (SINR) at the base-stations.  
This paper further studies the optimal
backhaul rate allocation problem for an uplink multicell joint
processing model with a total backhaul capacity constraint. The
analysis reveals that the optimal rate allocation that maximizes the
overall sum rate should have backhaul link rates that also scale 
logarithmically with the SINR at
each base-station. Finally, the proposed per-base-station SIC scheme
is evaluated in a practical multicell
orthogonal frequency division multiple access (OFDMA) network to
quantify the performance gain brought by the centralized processor. 
\end{abstract}


\section{Introduction}

Traditional wireless cellular networks operate on a cell-by-cell 
basis where out-of-cell interference is treated as part of the noise.
However, as cellular base-stations become more densely deployed and
heterogeneous networks become commonplace where small cells overlap
heavily with macro-cells, modern cellular networks are becoming
increasingly interference limited. Because of the intercell
interference, the per-cell achievable rates in traditional cellular
deployments are typically much smaller than that of a single isolated
cell.  

Multicell joint processing is a promising technique that has the
potential to significantly improve the cellular throughput by taking
advantage of its capability for intercell interference mitigation.
When base-stations share the transmitted and received signals, the
codebooks, and the channel state information (CSI) with each other through
high-capacity backhaul, it is theoretically possible to perform
joint transmission in the downlink and joint reception in the uplink
to eliminate out-of-cell interference completely. This paper deals
with the information theoretical capacity analysis of the
joint-processing architecture in the uplink. In this architecture,
antennas from multiple base-stations essentially become a virtual
multiple-input multiple-output (MIMO) array capable of spatial
multiplexing multiple user terminals. 

One way to implement a multicell joint processing system is to 
deploy a centralized processing server which is connected to all the
base-stations via high-capacity backhaul links. When the capacities of
the backhaul links are sufficiently large, the joint processing system
implemented across the different cells in the network can be
modeled as a multiple-access channel in the uplink and a broadcast
channel in the downlink, giving rise to the concept of network MIMO
\cite{network_mimo, new_look_at_interference}.

\begin{figure} [t]
\centering
\begin{overpic}[width=3.2in]{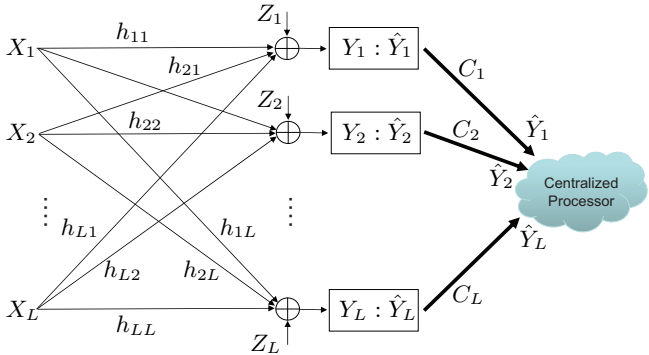}
\centering
\put(50,47.5){\small $Y_1:\hat{Y}_1$}
\put(50,33.5){\small $Y_2:\hat{Y}_2$}
\put(49.5,4.5){\small $Y_L:\hat{Y}_L$}
\put(-5,47.5){\small $X_1$}
\put(-5,33.5){\small $X_2$}
\put(-5,4.5){\small $X_L$}
\put(36,53){\small $Z_1$}
\put(36,39){\small $Z_2$}
\put(35,-1){\small $Z_L$}
\put(13,50){\small $h_{11}$}
\put(21.5,44.5){\small $h_{21}$}
\put(4,18){\small $h_{L1}$}
\put(30,18){\small $h_{1L}$}
\put(11,11){\small $h_{L2}$}
\put(24,11){\small $h_{2L}$}
\put(15,35.5){\small $h_{22}$}
\put(13,2){\small $h_{LL}$}
\put(69,44){\small $C_1$}
\put(68,34){\small $C_2$}
\put(68,7){\small $C_L$}
\put(80,34){\small $\hat{Y}_1$}
\put(74,26){\small $\hat{Y}_2$}
\put(79,16){\small $\hat{Y}_L$}
\end{overpic}
\caption{Uplink multicell joint processing via a centralized processor}
\label{cloud_model}
\end{figure}

The practical implementation of a network MIMO system must
also consider the effect of finite capacity in the backhaul. While the
capacity of the network MIMO system with infinite backhaul is easy to
compute, when finite backhaul is considered, the information
theoretical analysis of multicell processing becomes significantly
more complicated. This paper focuses on an uplink network MIMO model
as shown in Fig.~\ref{cloud_model}. From an information theoretical
perspective, the overall system can be thought of as a combination
of a multiple-access channel (with remote terminals acting as the
transmitters and the centralized processor as the receiver) and a
relay channel (with the base-stations acting as relays).

\subsection{Related Work}

The uplink network MIMO model considered in this paper, 
where the joint processing takes place in the central server, 
is related to but is different from the multicell joint processing model
with finite-capacity uni- or bi-direction backhaul links deployed
between the base-stations.  In this latter setting, the uplink multicell
model becomes an interference channel with partial receiver
cooperation. Although a complete characterization of capacity for such
a model is still an open problem, information theoretical studies have been
carried out for the case of two-user Gaussian interference channel
with receiver cooperation \cite{Wang_ReceiverCooperation,
Host-Madsen_cooperation, Jindal_cooperation,
Promond_DestinationCooperation, Skoglund_1}.
Likewise, the analysis of uplink cellular network under this setup has
been carried out in \cite{Marsch_uplink_CoMP, Osvaldo_localBS}, where
strategies such as
decode-and-forward and compress-and-forward are proposed, 
and analytical results under the Wyner model are obtained.
We mention briefly here that the downlink counterpart of this model
has been studied in \cite{Chowdhery}. 

For the uplink multicell joint processing model considered in this
paper, where the limited-capacity backhaul links are deployed between
the base-stations and a central processor as
depicted in Fig.~\ref{cloud_model}, although a complete characterization of its
information theoretical capacity region is also an open problem, 
much work has been done on finding its achievable rate
region and the gap to the outer bound \cite{Sanderovich_mcp,
Sanderovich_decentralize, sumrate_multicell_proc, delcoso, park, Park_JDD}.
The early work by Sanderovich et al. \cite{Sanderovich_mcp} considers
a joint decoding scheme in which the base-stations quantize the
received signals and forward them to the centralized processor, which
then performs \emph{joint decoding} of both the source messages and quantized
codewords.  Instead of joint decoding, \cite{park, delcoso} describe
schemes which \emph{successively} decode the quantization codewords
first, then the source messages. A comparison of successive decoding
vs. joint decoding is contained in \cite{Park_JDD}.  Alternatively,
base-stations can also \emph{decode} part of the messages of users in
their own cell, then forward the decoded data along with the remaining
part to the centralized processor 
\cite{Sanderovich_mcp, Sanderovich_decentralize, Sanderovich_scalinglaw}.  
The counterpart of these results for the
downlink with finite backhaul has been described in \cite{Osvaldo_dl_multicell}.
A review of some of these results is available in
\cite{new_look_at_interference}.

The uplink multicell joint processing model is actually a special case of 
a general multiple multicast relay network studied in 
\cite{Avestimehr_relaynetwork, Kim_noisy_network_coding}, for
which a generalization of quantize-and-forword can be shown to achieve
to within a constant gap to the information theoretical capacity of
the overall network. This is proved in \cite{Avestimehr_relaynetwork}
using a quantize-map-and-forward technique, and in
\cite{Kim_noisy_network_coding} using noisy network coding, both of
which require the joint decoding of the source messages and the
quantized codewords.
Thus in retrospect, the joint processing scheme of
\cite{Sanderovich_mcp} already achieves the approximate capacity of
the uplink finite-backhaul multicell model, provided that the
quantization noise level is appropriately chosen and joint decoding is done
at the central processor. 




It should be emphasized that joint decoding is challenging to 
implement. Not only is the network-wide CSI
required at the central processor, 
the computational complexity of joint decoding typically scales
exponentially with the total number of nodes in the network.
Moreover, even a mere evaluation of the achievable rate can be 
computationally prohibitive.  The achievable rate region of the
joint decoding scheme as presented in \cite[Proposition IV.1]{Sanderovich_mcp} 
involves $2^L-1$ rate constraints, each requiring a minimization of
$2^L$ terms, where $L$ is the number of users in the uplink multicell
model. To obtain analytical results, \cite{Sanderovich_mcp} and
likewise \cite{Osvaldo_localBS, somekh_sum_rate} assume certain
symmetry and obtain rate expressions under a Wyner cellular model.
But the rate expression for an arbitrary network remains difficult to
evaluate. On top of this, the achievable rate region needs to be
optimized over all choices of $L$ quantization noise levels, which is 
a difficult task, although recent progress has been made in
\cite{delcoso,Park_JDD,Yuhan}.

\subsection{Main Results}

To fully overcome the aforementioned difficulty yet still take full 
advantage of the centralized processing is expected to be challenging.  
This paper instead aims to design schemes with simple
receiver structures that result in computationally feasible
achievable rate regions for the uplink multicell joint processing
problem. Toward this end, this paper focuses on the uplink multicell
model with finite-capacity backhaul to the centralized processor, and
proposes suboptimal schemes based on \emph{successive
interference cancellation} (SIC).  Instead of performing joint
decoding of the source messages and quantization codewords
\cite{Sanderovich_mcp}, or successive decoding of all the quantization
codewords first, then the source messages \cite{delcoso,park}, this
paper applies the Wyner-Ziv compress-and-forward relaying scheme on a
per-base-station basis and performs single-user decoding with SIC at
the centralized processor.  The Wyner-Ziv coding here also uses
the previously decoded messages as side information. 
The main advantage of this
proposed scheme is that it leads to a receiver architecture in which
only the decoded transmit signals are shared between the successive stages, so it
is more amenable to implementation. Further, the quantization noise levels
can now be easily optimized, resulting in an achievable rate region
computable with complexity that scales linearly with $L$.

Although the proposed per-base-station SIC scheme is no longer the
best possible for a general uplink joint processing model, this paper
shows that for a special Wyner soft-handoff model 
\cite{Wyner_softhandoff}, in which the mobile terminals and the
base-stations are placed along a line, the proposed per-base-station 
SIC scheme achieves a sum rate that is at most a constant gap away 
from capacity. This paper shows that the constant gap is a linear 
function of the number of transmitter-receiver pairs in the uplink 
multicell joint processing model.

Under the proposed per-base-station SIC framework for the
finite-backhaul uplink model, we also ask the following question: 
How much backhaul capacity is needed to approach the theoretical
infinite-backhaul successive-decoding rate?  The results of this paper
show that the backhaul rates need to scale logarithmically with the
received signal-to-interference-and-noise ratio (SINR) in order to
achieve to within $\frac{1}{2}$ bit of the successive decoding rates
attainable with unlimited backhaul capacity.

Further, this paper addresses the question of how to allocate the backhaul
rates optimally across the different backhaul links.  Under a sum
rate constraint on the total backhaul capacity, this paper shows
that in order to maximize the sum rate over the entire network, the
allocation of individual backhaul link capacity should again scale
logarithmically with the SINR under the proposed per-base-station SIC
framework. 

This paper focuses on capacity analysis with the assumption that the
CSI for the entire network is available at the central processor. This
is reasonable as the base-stations may estimate the channel in the
uplink, then send CSI via the backhaul links to the central processor.
For convenience, this paper neglects the additional backhaul capacity
needed to support the sharing of CSI.  We mention that the capacity of
network MIMO under channel estimation error has been investigated in
\cite{Marsch_uplink_CoMP}.  Study on the use of processing techniques
that are robust against channel uncertainty has been reported in
\cite{park}. 

Finally, under the perfect CSI assumption, this paper evaluates the
performance of the proposed per-base-station SIC scheme in an OFDMA 
network. 
Numerical simulations show that
the proposed per-base-station SIC scheme can almost double 
the per-cell sum-rate over a baseline scheme where no centralized
processor is deployed and each base-station simply treats interference
as noise.  Further, with optimized backhaul capacity allocation,
most of such gains can already be obtained when the
backhaul sum capacity is at about 1.5 times the achieved per-cell sum
rate.

\subsection{Organization of the Paper}

The rest of the paper is organized as follows. Section II presents 
the multicell joint processing model. Section III introduces the
per-base-station SIC scheme and compares it with other achievability
schemes such as noisy network coding and joint-base-station
processing. The optimality of the per-base-station SIC scheme for the
Wyner model is provided in Section IV. The backhaul rate allocation
problem with a total backhaul constraint is solved in Section V.
Finally, Section VI numerically evaluates the per-base-station SIC
scheme for a two-user symmetric setting and for a realistic OFDMA
network.


\section{Channel Model}
%

Consider the uplink of a multicell network with a central processor.
Assuming that there is only one user operating in each time-frequency
resource block in each cell, the multicell network can be modelled by $L$
users each sending a message to their corresponding  base-stations.
Base-stations serve as intermediate relays for the
centralized server, which eventually decodes all the transmit
messages.  

As depicted in Fig.~\ref{cloud_model}, the uplink joint processing
model consists of two parts. The left half is an $L$-user interference
channel with $X_i$ as the input signal from the $i$th user, $Y_i$
as the output signal, $Z_i$ as the additive white Gaussian noise
(AWGN), and $h_{ij}$ as the channel gain from user $i$ to base-station $j$,
where $i,j =1,2, \cdots, L$. The right half can be seen as a
noiseless multiple-access channel with capacities $C_i$, 
$i=1, 2, \cdots, L$, modelling the backhaul links 
between the base-stations and the central processor.
The input signal $X_i$ is power constrained, i.e., 
$E[|X_i|^2] \le P_i$; the receiver noises are assumed to be independent 
and identically distributed with $Z_i \sim \mathcal{N}(0, N_0)$, 
$i =1,2, \cdots, L$. The signal-to-noise ratios (SNRs) and the 
interference-to-noise
ratios (INRs) are defined as follows: 
\begin{equation}
\mathsf{SNR}_i = \frac{h_{ii}^2P_i}{N_0}, \quad \mathsf{INR}_{i,j} = \frac{h_{ij}^2P_i}{N_0}, \quad i, j = 1, \cdots, L
\end{equation}
In addition, define the vectors $\mathbf{X} = \{ X_1, X_2, \cdots, X_L\}$ 
and $\mathbf{Y} =\{Y_1, Y_2, \cdots, Y_L\}$.


\section{Multicell Processing Schemes}

This paper focuses on uplink multicell processing schemes in which 
the base-stations quantize the received signals, then subsequently
transmit a function of the quantized signal to the central processor
via noiseless backhaul links. Quantize-and-forward is a natural
strategy in the uplink setting. This is because in order to reap the benefit of
multicell processing, the user terminals must transmit at rates 
higher than that decodable at its own base-station alone. This prevents
the use of the decode-and-forward strategy. In Fig.~\ref{cloud_model},
the received signal is denoted as $Y_i$; its quantized version is 
denoted as $\hat{Y}_i$. This paper also assumes that the base-stations
either transmit $\hat{Y}_i$ directly, or performs a binning
operation on $\hat{Y_i}$. The task at the central processor is to
decode $\{X_1,X_2,\cdots,X_L \}$ based on either $\{\hat{Y_1},
\hat{Y_2}, \cdots, \hat{Y_L}\}$ or their bin indices. In the
following, we examine the information theoretical achievable rates for
various quantization and decoding strategies.

\subsection{Joint Decoding}

Consider a coding scheme in which each base-station quantizes 
its received signal using a Gaussian codebook at certain distortion or
quantization noise level $\mathsf{q}_i$, bins the quantized message, 
then forwards the bin index to the central processor.  In the joint decoding
strategy, the central processor, upon receiving all the bin indices, decodes
$\{X_1,X_2,\cdots,X_L\}$ and $\{\hat{Y}_1,\hat{Y}_2,\cdots,\hat{Y}_L\}$ jointly
based on joint typicality.  This strategy is first proposed by Sanderovich et
al.\ \cite[Proposition IV.1]{Sanderovich_mcp}. (Note that there is no
requirement that $\{\hat{Y}_1,\hat{Y}_2,\cdots,\hat{Y}_L\}$ are decoded
correctly; only $\{X_1,X_2,\cdots,X_L\}$ need to.)

The key parameter here is the setting of the quantization noise levels
$\mathsf{q}_i$.  For example, $\mathsf{q}_i$'s can be set to be such that
$\{\hat{Y}_1,\hat{Y}_2,\cdots,\hat{Y}_L\}$ can be correctly decoded based 
on the bin indices. The values of $\mathsf{q}_i$'s can then be determined 
based on the capacities of the backhaul links $(C_1,C_2,\cdots,C_L)$.
But, such a setting of $\mathsf{q}_i$ may not be optimal. 
Indeed, as shown in \cite{Avestimehr_relaynetwork} and 
also later in \cite{Kim_noisy_network_coding}, for a much larger class
of general relay networks, setting the quantization noise levels to be at
the background noise level, i.e., $\mathsf{q}_i = N_0$, leads to an achievable rate
that is within at most a constant gap from the capacity outer bound.
This gap scales linearly with the number of nodes in the
network, but is independent of the channel gains and the backhaul
capacities.  Applying this result to the uplink multicell model in
Fig.~\ref{cloud_model}, we immediately have the following joint
decoding achievable rate.  The achievable rate expression can be
derived based on the noisy-network-coding theorem
\cite{Kim_noisy_network_coding}, and is equivalent to the expression
in \cite{Sanderovich_mcp}.

\begin{lemma} \label{lema:nnc_region}
For the multicell joint processing model as shown in
Fig.~\ref{cloud_model}, the following rate region is achievable 
\begin{multline}
R(\mathcal{S}) \le 
 \min_{\mathcal{T} \subseteq \mathcal{L}} \left \{ \frac{1}{2} \log
 \left | I +  \Lambda^{-1}_{\mathcal{\mathcal{T}}^c}(N_0 +\mathsf{q})
    \mathbf{H}_{\mathcal{ST}^{c}}  \Lambda_{\mathcal{S}}(P)
    \mathbf{H}_{\mathcal{ST}^{c}}^{T} \right| \right. \\ 
+ \left.  \sum_{i \in \mathcal{T}} \left(C_i - \frac{1}{2} \log (1 +
    \frac{N_0}{\mathsf{q}_i}) \right) \right\}, \quad 
 \forall \; \mathcal{S} \subseteq \mathcal{L} \triangleq 
	\{0, 1, \cdots, L\}   \label{rate_NNC} 
\end{multline} 
where $R(\mathcal{S}) = \sum_{i \in \mathcal{S}} R_i$,
$\mathsf{q}_i$ is a positive real number representing the quantization
noise level at base-station $i$, 
$\mathbf{H}_{\mathcal{ST}^{c}}$ denotes for the transfer
matrix from input $\mathbf{X}(\mathcal{S})$ to output
$\mathbf{Y}(\mathcal{T}^c)$,
$\Lambda^{-1}_{\mathcal{\mathcal{T}}^c}(N_0 +\mathsf{q}) $ is a
diagonal matrix of $\frac{1}{N_0 + \mathsf{q}_i}$ with $i \in
\mathcal{T}^c$, and $\Lambda_{\mathcal{S}}(P)$ is a diagonal matrix of
$P_i$ with $i \in \mathcal{S}$.  
This rate region is within a constant gap to the capacity region
of the multicell joint processing model if we set
$\mathsf{q}_i = N_0, i = 1, 2, \cdots, L$. 
\end{lemma}
\vspace{1em}

The above achievable rate region is valid for any choice of
$\mathsf{q}_i$, and is approximately optimal when $\mathsf{q}_i=N_0$.
But, the evaluation of such an achievable rate region can be
difficult. For the uplink joint processing model, the achievable 
rate region as derived in Lemma~\ref{lema:nnc_region} requires a
minimization of $2^L$ terms for each rate constraint, and there are
$2^L-1$ different rate constraints describing the rate region. Even
when the size of the network is in a reasonable range, for example as
in a $19$-cell topology, it is computationally prohibitive to minimize
over $2^{19}$ terms each involving $2^{19}-1$ different rate
constraints.  Further, the implementation of the joint decoding scheme
itself tends to have exponential complexity, so the achievable rate
given in Lemma~\ref{lema:nnc_region} is not practically implementable
for a reasonably-sized network.

In order to derive more tractable performance analysis of the
multicell joint processing scheme, \cite{Sanderovich_mcp} resorts to 
a modified Wyner model (see \cite{Wyner_softhandoff}), where each
transmitter-receiver pair interferes only with one other neighboring
transmitter-receiver pair, and is subject to interference from only
one neighboring transmitter-receiver pair. Further, certain symmetry
is introduced so that all the direct channels are identical, and so
are all interfering channels. With this symmetric cyclic structure, 
computation of the sum rate becomes tractable under joint decoding 
\cite{Sanderovich_mcp}.

\subsection{Per-Base-Station SIC}

This paper focuses on the general multicell model (instead of the
symmetric Wyner model) and proposes suboptimal schemes
based on the successive decoding of source messages. 
Based on the observation that the exponential complexity of the rate
expression in Lemma~\ref{lema:nnc_region} is due to the joint decoding
step at the destination, this paper proposes a per-base-station
SIC approach (in contrast to joint-base-station SIC
to be described in detail later) at the centralized processor that
significantly reduces the complexity of computation of the
achievable rate region. In addition, the proposed scheme involves the
use of compress-and-forward relaying technique \cite{Cover1979} at each
base-station independently, which also significantly reduces the
complexity of its eventual implementation in a practical system.

Specifically, we use an SIC approach assuming a \emph{fixed} order of
decoding first $X_1$, then $X_2, X_3, \cdots, X_L$.  A central feature
of the proposed scheme is that only the decoded signals are used as
side information in subsequent stages. This simplifies the sharing
of information among the base-stations and facilitates practical
implementation.

At the $k${th} stage, 
the base-station $k$, upon receiving $Y_k$, quantizes $Y_k$ into
$\hat{Y}_k$ using the compress-and-forward technique and sends bin
index of $\hat{Y}_k$ to the destination via the noiseless link of
capacity $C_k$.  Note that the quantization process at the
base-station $k$ includes interference from all other users.  The
quantization noise level at each base-station $k$ is chosen so as to
fully utilize $C_k$ such that $\hat{Y}_k$ can be immediately recovered
at the central processor.  This quantization process can be done
either with Wyner-Ziv coding that takes advantage of the
already-decoded user messages $(X_1,\cdots, X_{k-1})$ as side
information, or without Wyner-Ziv coding for ease of implementation.  

At the central processor, to decode user $k$'s message $X_k$ the
central processor first decodes the quantization message $\hat{Y}_k$
upon receiving its description from the digital link $C_k$. It then
decodes the message of user $k$ using joint typicality decoding between 
$X_k$ and the quantized message $\hat{Y}_k$. The decoding of $X_k$ also
takes advantage of the knowledge of previously decoded messages $(X_1,
\cdots, X_{k-1})$ at the centralized processor. In this way, the
impact of interference from $X_1, \cdots, X_{k-1}$ eventually
disappears and the effective interference is only due to users not yet
decoded, i.e., $X_j$, for $j > k$. After decoding $X_k$, the central
processor moves to the next decoding stage, adding $X_k$ to the set
of known side information.

\subsubsection{Per-Base-Station SIC with Wyner-Ziv}
The following theorem gives the achievable rate of the proposed
per-base-station SIC scheme with Wyner-Ziv compress-and-forward
relaying.

\begin{theo} \label{theo: per-base-station-SIC-WZ}
For the uplink multicell joint processing channel depicted in
Fig.~\ref{cloud_model}, the following rate is achievable using
Wyner-Ziv compress-and-forward relaying at the base-stations followed
by SIC at the centralized processor
with a fixed decoding order:
\begin{equation} \label{rate:per_BS_SIC_WZ}
R_k = \frac{1}{2} \log \frac{1 + \mathsf{\overline{SINR}}_k}{1 + 2^{-2C_k}\mathsf{\overline{SINR}}_k},
\end{equation}
where
\begin{equation} \label{SINRk}
\mathsf{\overline{SINR}}_k = \frac{\mathsf{SNR}_k}
{1 + \sum_{j > k} \mathsf{INR}_{j,k}}.
\end{equation}
\end{theo}


\begin{figure} [t]
\centering
\vspace{1.2em}
\begin{overpic}[width=3.2in]{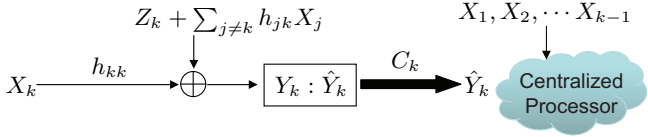}
\centering
\put(-5,7){\small $X_k$}
\put(9,10){\small $h_{kk}$}
\put(58,11){\small $C_{k}$}
\put(16,18){\small $Z_k + \sum_{j \neq k}h_{jk}X_j$}
\put(39,7){\small $Y_k : \hat{Y}_k$}
\put(70,7){\small $\hat{Y}_k$}
\put(69,19){\small $X_1, X_2, \cdots X_{k-1}$}
\end{overpic}
\caption{Equivalent channel of user $k$ in the $k^{th}$ decoding stage}
\label{CF_user_k}
\end{figure}

\begin{IEEEproof}
In the $k${th} stage of the SIC
decoder, $X_1, \cdots, X_{k-1}$ decoded in the previous decoding stages
serve as side information for stage $k$. The equivalent channel of
user $k$ is depicted in Fig.~\ref{CF_user_k}.  This is a three-node
relay channel without the direct source-destination link.
Specifically, the source signal $X_k$ is sent from the transmitter to the
relay, which receives $Y_k$, quantizes it into $\hat{Y}_k$ and forwards
its description to the centralized processor via the noiseless
digital link of capacity $C_k$. At the centralized processor, $X_1, \cdots, X_{k-1}$ serve as side information and facilitate the decoding of $\hat{Y}_k$ and $X_k$. According to \cite[Theorem 6]{Cover1979}, the achievable rate of user $k$ using Wyner-Ziv compress-and-forward can be written as
\begin{equation}
R_k = I(X_k; \hat{Y}_k | X_1, \cdots, X_{k-1})
\end{equation}
subject to the constraint
\begin{equation}
I(Y_k; \hat{Y}_k | X_1, \cdots, X_{k-1}) \le C_k.
\end{equation}

We constrain ourselves to Gaussian input signals $X_k \sim \mathcal{N}(0, P_k)$, 
and the Gaussian quantization scheme\footnote{Gaussian input and Gaussian 
quantization are used for convenience and for simplicity only. They are not 
necessarily optimal; see \cite{Sanderovich_decentralize} for an example.}, 
i.e., 
\begin{equation}
\hat{Y}_k = Y_k + e_k,
\end{equation}
where $e_k \sim \mathcal{N}(0, \q_k)$ is the Gaussian quantization noise 
independent of everything else.
To fully utilize the digital link, it is natural to set
\begin{equation} \label{backhaul_utilization}
I(Y_k; \hat{Y}_k | X_1, \cdots, X_{k-1}) = C_k.
\end{equation}
Now, substituting $Y_k = \sum_{j=1}^{L}h_{jk}X_j + Z_k$ and $\hat{Y}_k = Y_k + e_k$ into (\ref{backhaul_utilization}), we have
\begin{eqnarray}
C_k &=& I(Y_k; \hat{Y}_k | X_1, \cdots, X_{k-1}) \nonumber \\
&=&h(\hat{Y}_k | X_1, \cdots, X_{k-1}) - h(e_k)  \nonumber \\
&=& \frac{1}{2} \log \left( 1 + \frac{N_0 + \sum_{j \ge k}h_{jk}^2 P_j}{\q_k}\right),
\label{Ck}
\end{eqnarray}
which results in the following quantization noise level that fully utilizes the digital links $C_k$:
\begin{equation} \label{q}
\q_k^* = \frac{N_0 + \sum_{j \ge k}h_{jk}^2 P_j}{2^{2C_k} - 1}.
\end{equation}
With the above $\q_k^*$, the achievable rate of user $k$ can be calculated as
\begin{eqnarray}
R_k &=& I(X_k; \hat{Y}_k | X_1, \cdots, X_{k-1}) \nonumber \\
&=& h( \hat{Y}_k | X_1, \cdots, X_{k-1}) - h( \hat{Y}_k | X_1, \cdots, X_{k}) \nonumber \\
&=& \frac{1}{2} \log \frac{\q_k^* + N_0 + \sum_{j \ge k}h_{jk}^2P_j }{\q_k^* + N_0 + \sum_{j > k}h_{jk}^2P_j } \nonumber \\
&=& \frac{1}{2} \log \frac{N_0 + \sum_{j > k}h_{jk}^2P_j  + h_{kk}^2 P_k}{N_0 + \sum_{j  > k}h_{jk}^2P_j   + 2^{-2C_k}h_{kk}^2P_k} \nonumber \\
&=&\frac{1}{2} \log \frac{1 + \mathsf{\overline{SINR}}_k}{1 + 2^{-2C_k}\mathsf{\overline{SINR}}_k}.
\label{Rk}
\end{eqnarray}
Finally, note that 
$
\mathsf{\overline{SINR}}_k = \frac{h_{kk}^2 P_k}{N_0 + \sum_{j > k}h_{jk}^2 P_j},
$
which is equivalent to (\ref{SINRk}).
\end{IEEEproof}
\vspace{1em}

\begin{figure} [t]
\vspace{0.5em}
\centering
\begin{overpic}[width=3.2in]{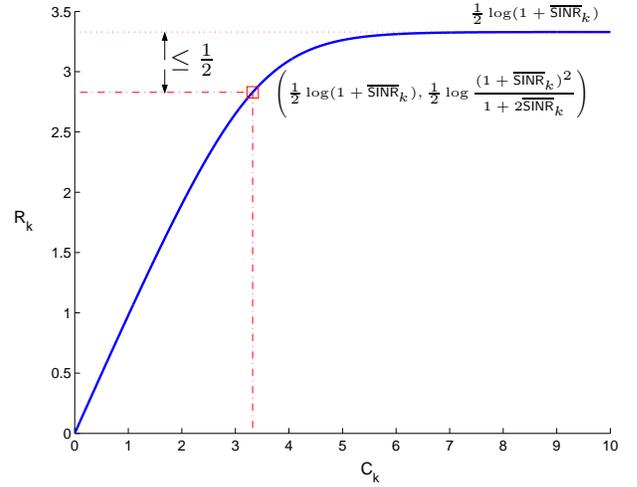}
\centering
\put(75,77){\tiny $\frac{1}{2}\log (1 + \mathsf{\overline{SINR}}_k)$}
\put(43,64){\tiny $\left(\frac{1}{2}\log (1 + \mathsf{\overline{SINR}}_k), \frac{1}{2} \log \displaystyle \frac{(1+\mathsf{\overline{SINR}}_k)^2}{1 + 2 \mathsf{\overline{SINR}}_k}  \right)$}
\put(26,69){$\le \frac{1}{2}$}
\end{overpic}
\caption{Achievable rate of user $k$ versus the backhaul capacity $C_k$}
\label{Rk_vs_Ck}
\end{figure}

The rate expression (\ref{rate:per_BS_SIC_WZ}) shows how the achievable
rates are affected by the limited capacities of the digital backhaul links
under the proposed per-base-station SIC decoding framework.
Fig.~\ref{Rk_vs_Ck} plots the achievable rate of $R_k$ as a function
of the backhaul link capacity $C_k$ with $\mathsf{\overline{SINR}}_k$ equal to $20$dB.
When $C_k$ is small, $R_k$
grows almost linearly with $C_k$, which means that each bit of the
backhaul link provides approximately one bit increase in the achievable rate
for user $k$. The digital backhaul is efficiently exploited in this
regime. However, as
$C_k$ grows larger, each bit of the backhaul link gives increasingly
less achievable rate improvement.  In the limit of unbounded backhaul capacity,
i.e., $C_k = \infty$,  $R_k$ saturates and approaches 
$ \frac{1}{2} \log (1 + \mathsf{\overline{SINR}}_k)
\stackrel{\triangle}{=} \bar{R}_k$,  which is 
the upper limit for the rate of user $k$ when the SIC decoder is employed.

Since backhaul link capacity can be costly in practical implementations, 
it is natural to ask how large $C_k$ needs to be in order to
achieve a rate $R_k$ that is close to the maximum SIC rate with
unlimited backhaul.
It is easy to see that when $C_k = \frac{1}{2} \log (1 + \mathsf{\overline{SINR}}_k)$,
$\bar{R}_k - R_k$ is upper bounded by one-half bit, i.e.,
\begin{eqnarray}
\bar{R}_k - R_k & =& \frac{1}{2} \log (1 + \mathsf{\overline{SINR}}_k)  \nonumber \\
&& -  \left. \frac{1}{2} \log \frac{1 + \mathsf{\overline{SINR}}_k}{1 + 2^{-2C_k}\mathsf{\overline{SINR}}_k} \right|_{C_{k} = \frac{1}{2}\log(1 + \mathsf{\overline{SINR}}_k)} \nonumber \\
&=& \frac{1}{2} \log \left(1 + \frac{\mathsf{\overline{SINR}}_k}{1 + \mathsf{\overline{SINR}}_k} \right) \nonumber \\
&\le& \frac{1}{2}.
\end{eqnarray}
Therefore, when the backhaul link has capacity $C_k = \frac{1}{2} \log (1 +
\mathsf{\overline{SINR}}_k)$, the achievable rate is half a bit away from the
SIC upper limit.  This is also the point under which the utilization
of $C_k$ is the most efficient, as shown in Fig.~\ref{Rk_vs_Ck}.

It is worth noting that the above result, which suggests that $C_k$ should
scale as $\log(\mathsf{\overline{SINR}}_k)$, is analogous to the conclusion
of \cite[Corollary IV.6]{Sanderovich_mcp}, which states that the backhaul rate
should scale as $\log(P)$ in order to approach the capacity of the
infinite backhaul case for the symmetric Wyner model. The main difference is
that this paper uses SIC, so interference not yet cancelled still appears
in the SINR expression.

\subsubsection{Per-Base-Station SIC Without Wyner-Ziv}

Implementing the Wyner-Ziv compression may not be easy in practice. 
The main difficulty lies in the binning operation that is necessary in
order to take advantage of the side information. To make the proposed 
per-base-station SIC scheme more amenable to practical implementation,
the Wyner-Ziv quantizer can be replaced by a simple vector
quantizer that does not take side information (i.e., previously
decoded messages in the SIC framework) into account. In this way, the
following rate can be achieved for user $k$: 
\begin{equation}
R_k = I(X_k; \hat{Y}_k | X_1, \cdots, X_{k-1})
\end{equation}
subject to the constraint
\begin{equation}
I(Y_k; \hat{Y}_k ) \le C_k,
\end{equation}
for $k=1, 2, \cdots, L$. Following the same lines of the proof of
Theorem~\ref{theo: per-base-station-SIC-WZ}, it is easy to see that
by replacing (\ref{Ck}) with
$C_k = \frac{1}{2} \log \left( 1 + \frac{N_0 + \sum_{j}h_{jk}^2 P_j}{\q_k}\right)$,
which gives $\q_k^* = \frac{N_0 + \sum_{j}h_{jk}^2 P_j}{2^{2C_k} - 1}$, 
we obtain
\begin{align}
R_k 
&= \frac{1}{2} \log \frac{\q_k^* + N_0 + \sum_{j \ge k}h_{jk}^2P_j }{\q_k^* + N_0 + \sum_{j > k}h_{jk}^2P_j } \nonumber \\
&= \frac{1}{2} \log \frac{N_0 + \sum_{j \ge k}h_{jk}^2P_j  + 
2^{-2C_k} \sum_{j < k} h_{jk}^2P_j}
{N_0 + \sum_{j> k}h_{jk}^2P_j + 2^{-2C_k}\sum_{j\ge k} h_{jk}^2P_j}.
\end{align}
Define 
\begin{equation}
\mathsf{\overline{SINR}'}_k = \frac{h_{kk}^2 P_k}{N_0 + \sum_{j > k}h_{jk}^2 P_j + 2^{-2C_k}\sum_{j<k}h_{jk}^2P_j}.
\end{equation}
We have just proved that the following rate is achievable using
per-base-station SIC without Wyner-Ziv compress-and-forward.

\begin{theo} \label{theo: per-base-station-SIC-NOWZ}
For the uplink multicell joint processing channel depicted in
Fig.~\ref{cloud_model}, the following rate is achievable using
vector quantization at the base-stations followed
by SIC at the centralized processor
with a fixed decoding order:
\begin{equation} \label{rate:per_BS_SIC_NOWZ}
R_k = \frac{1}{2} \log \frac{1 + \mathsf{\overline{SINR}'}_k}{1 +
2^{-2C_k}\mathsf{\overline{SINR}'}_k},
\end{equation}
where
\begin{equation}\label{SINR: per_BS_SIC_NOWZ}
\mathsf{\overline{SINR}'}_k = \frac{\mathsf{SNR}_k}
{1 + \sum_{j>k} \mathsf{SIR}_{j,k} + 2^{-2C_k} \sum_{j<k} \mathsf{SIR}_{j,k}}.
\end{equation}
\end{theo}
\vspace{1em}

Comparing Theorem~\ref{theo: per-base-station-SIC-NOWZ} with
Theorem~\ref{theo: per-base-station-SIC-WZ}, it is easy to see that
achievable rates in (\ref{rate:per_BS_SIC_WZ}) and
(\ref{rate:per_BS_SIC_NOWZ}) share the same structure. However,
without Wyner-Ziv compression, the effective SINR
(\ref{SINR: per_BS_SIC_NOWZ}) 
becomes smaller due to the extra terms $2^{-2C_k}\sum_{j<k}
\mathsf{SIR}_{j,k}$. This means that the impact of previously decoded signals
$X_1, \cdots, X_{k-1}$ is not fully cancelled at stage $k$.
We also note that as $C_k \rightarrow \infty$, we have
$ \mathsf{\overline{SINR}'}_k \rightarrow 
\mathsf{\overline{SINR}}_k$. Thus, Wyner-Ziv coding is most useful
only when the backhaul link capacity is small. For reasonably large
$C_k$, the benefit of Wyner-Ziv coding is expected to be marginal.

\subsection{Joint Base-Station SIC Schemes}
\label{joint-bs-processing}

The per-base-station SIC scheme takes advantage of multicell
processing only in so far as to the use of the decoded messages as
side information. It is possible to further improve upon these schemes
by joint decoding across the base-stations. 

Note that in the 
per-base-station SIC scheme, the decoding of source messages and
quantized messages follow the order $\hat{Y}_1 \rightarrow X_1
\rightarrow \hat{Y}_2 \rightarrow X_2, \rightarrow \cdots, \rightarrow
\hat{Y}_{L} \rightarrow X_L$, where in the $k$th compression and
decoding stage, previously decoded source signals $X_1, \cdots,
X_{k-1}$ are used as side information. However, note that in the
decoding process the quantization codewords $\hat{Y}_1, \cdots, 
\hat{Y}_{k-1}$ are also decoded along the way, and thus are also 
available for possible joint processing at stage $k$, which can lead
to better performance.  In particular, we can incorporate 
$\hat{Y}_1, \cdots, \hat{Y}_{k-1}$ in the expressions for $C_k$ and
$R_k$, leading to the following achievable rate: 
\begin{equation} \label{rate: improved-per-base-station-SIC}
R_k = I(X_k; \hat{Y}_1, \cdots, \hat{Y}_{k} | X_1, \cdots, X_{k-1}),
\end{equation}
subject to
\begin{equation}
C_k \ge I(Y_k: \hat{Y}_k|X_1, \cdots, X_{k-1}, \hat{Y}_1, \cdots, \hat{Y}_{k-1}),
\end{equation}
for $k=1, 2, \cdots, L$. 

Alternatively, we can also proceed in a
two-stage successive process of decoding all of $\{ \hat{Y}_k \}_{k=1}^L$ first,
then $\{X_k \}_{k=1}^L$ \cite{park,delcoso}. Further, each of these
two stages can be accomplished in an SIC fashion, resulting in rate
expressions 
\begin{equation} \label{rate: joint-BS SIC}
R_k = I(X_k; \hat{Y}_1, \cdots, \hat{Y}_L | X_1, \cdots, X_{k-1}),
\end{equation}
subject to
\begin{equation}
I(Y_k; \hat{Y}_k | \hat{Y}_1, \cdots, \hat{Y}_{k-1}) \le C_k,
\label{park_1}
\end{equation}
for $k=1, 2, \cdots, L$.
We note here that the above rate expressions (\ref{rate:
improved-per-base-station-SIC}) and (\ref{rate: joint-BS SIC}) can
outperform the rates (\ref{rate:per_BS_SIC_WZ}) in
Theorem~\ref{theo: per-base-station-SIC-WZ} and
(\ref{rate:per_BS_SIC_NOWZ}) in Theorem~\ref{theo:
per-base-station-SIC-NOWZ}, because in the above expressions each
$X_k$ is decoded based on the quantized observations of all
base-stations (or that of all previous decoded base-stations),
rather than just the $k$th base-station in the per-base-station SIC
scheme. However, for this same reason, the implementations of the above 
joint-base-station SIC schemes are also expected to be somewhat more complicated. 
For the rest of this paper, we only focus on the per-base-station SIC
schemes of Theorem~\ref{theo: per-base-station-SIC-WZ} and
Theorem~\ref{theo: per-base-station-SIC-NOWZ}, and leave the
joint-base-station SIC schemes as subject for future studies.

\section{Approximate Optimality of Per-Base-Station SIC in a Wyner Model}

\begin{figure} [t]
\vspace{1em}
\centering
\begin{overpic}[width=3.2in]{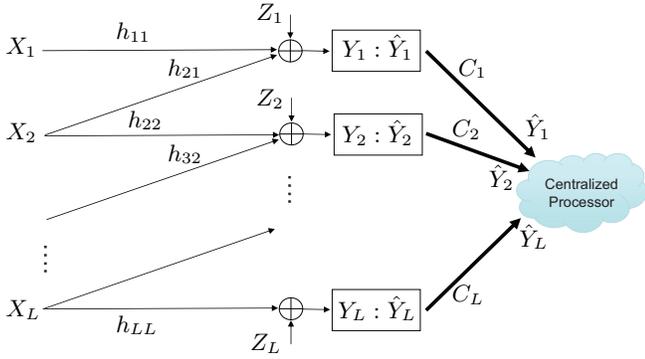}
\centering
\put(50,47){\small $Y_1:\hat{Y}_1$}
\put(50,33){\small $Y_2:\hat{Y}_2$}
\put(49.3,4.3){\small $Y_L:\hat{Y}_L$}
\put(-5,47.5){\small $X_1$}
\put(-5,33.5){\small $X_2$}
\put(-5,4.5){\small $X_L$}
\put(36,53){\small $Z_1$}
\put(36,39){\small $Z_2$}
\put(35,-1){\small $Z_L$}
\put(13,50){\small $h_{11}$}
\put(21.5,44){\small $h_{21}$}
\put(21.5,30){\small $h_{32}$}
\put(15,35.5){\small $h_{22}$}
\put(13,2){\small $h_{LL}$}
\put(69,44){\small $C_1$}
\put(68,34){\small $C_2$}
\put(68,7){\small $C_L$}
\put(80,34){\small $\hat{Y}_1$}
\put(74,25.7){\small $\hat{Y}_2$}
\put(79,16){\small $\hat{Y}_L$}
\end{overpic}
\caption{Wyner model where each transmitter-receiver only interferes 
with one neighbor.}
\label{asymmetric}
\end{figure}

The proposed per-base-station SIC scheme is in general sub-optimal.
However, in certain asymmetric settings, the per-base-station SIC
scheme can be shown to be approximately optimal in the sense of
achieving the sum capacity to within a constant gap. This section
studies a class of such asymmetric channels in which each
transmitter-receiver pair interferes only with one neighbor and gets
interfered by only one other neighbor as shown in Fig.~\ref{asymmetric}.  
This model is called the soft-handoff Wyner model in the literature
\cite{Wyner_softhandoff}.  The Wyner model is an abstraction of the
cellular network to a one-dimensional setting. It can, for example,
model a communication scenario where base-stations are placed along
a highway and mobile terminals travel between the base-stations. This
section proves that the per-base-station SIC scheme achieves to within
a constant gap to capacity for the Wyner model. This result holds
either with or without Wyner-Ziv coding.

Consider first the per-base-station SIC with Wyner-Ziv coding.
Intuitively, decoding order plays an important role in this asymmetric
setting. Since receiver $L$ sees a non-interfered version of the source
signal $X_L$, decoding should take place starting from user $L$.
After $X_L$ is decoded, it can serve as side information, which 
facilitates the decoding of user $L-1$, then $L-2$, etc. With this
decoding order and applying the result of Theorem~\ref{theo:
per-base-station-SIC-WZ} to the channel setting of Fig.~\ref{asymmetric},
we arrive at the following achievable rate region: 
\begin{equation}
\label{achievable_rate_asymmetric}
\left\{
  \begin{array}{l}
R_i = I(X_i; \hat{Y}_i|X_{i+1}), \;\; i= 1, 2, \cdots, L-1 \\
R_L = I(X_L; \hat{Y}_L)
  \end{array}
\right.
\end{equation}
subject to
\begin{equation} \label{quantization:Wyner_model}
\left\{
  \begin{array}{l}
I(Y_i;\hat{Y}_i|X_{i+1}) \le C_i,  \;\; i= 1, 2, \cdots, L-1 \\
I(Y_L;\hat{Y}_L) \le C_L
  \end{array}
\right.
\end{equation}
which, when specialized to Gaussian inputs $X_i \sim \mathcal{N}(0, P_i)$ 
and the Gaussian quantization scheme: $\hat{Y}_i = Y_i + e_i$, where 
$e_i$ is the quantization noise independent of everything else and
follows
Gaussian distributions $e_i \sim \mathcal{N}(0, \mathsf{q}_i)$, 
gives the following achievable rate 
\begin{equation}
\label{rate_asymmetric_Gaussian}
R_i =  \D \frac{1}{2} \log \left ( \D \frac{1 + \mathsf{SNR}_i}{1 + 2^{-2C_i}\mathsf{SNR}_i}\right), \;\; i=1, 2, \cdots, L.
\end{equation}
The achievable sum rate for the Wyner model with decoding order from 
user $L$ to user $1$ is then 
\begin{equation}\label{sum_rate_asymmetric}
R_{\mathrm{sum}} = \sum_{i=1}^{L}\frac{1}{2} \log \left ( \D \frac{1 + \mathsf{SNR}_i}{1 + 2^{-2C_i}\mathsf{SNR}_i}\right).
\end{equation}
The above sum rate can be slightly improved by noticing that the $L$th 
base-station can use decode-and-forward instead of
compress-and-forward, which leads to 
\begin{equation}\label{RL_improved}
R_L = \min \left\{\frac{1}{2} \log(1+\mathsf{SNR}_L), C_L \right\}.
\end{equation}
The main result of this section is that under a mild weak interference 
condition, the sum rate achieved by
the per-base-station SIC scheme is at most a constant number of bits away
from the sum capacity for the Wyner model, so it is approximately
optimal in the high SNR regime.

\begin{theo}\label{theo:asymmetric}
For a multicell processing Wyner model shown in
Fig.~\ref{asymmetric}, in the weak-interference regime of
$\mathsf{INR}_{i+1, i} \le \mathsf{SNR}_i, i=1, 2, \cdots, L-1$, the
per-base-station SIC scheme with Wyner-Ziv coding achieves a sum rate
that is within $L-\frac{1}{2}$ bits of the sum capacity.
\end{theo}

\begin{figure} [t]
\centering
\begin{overpic}[width=3.2in]{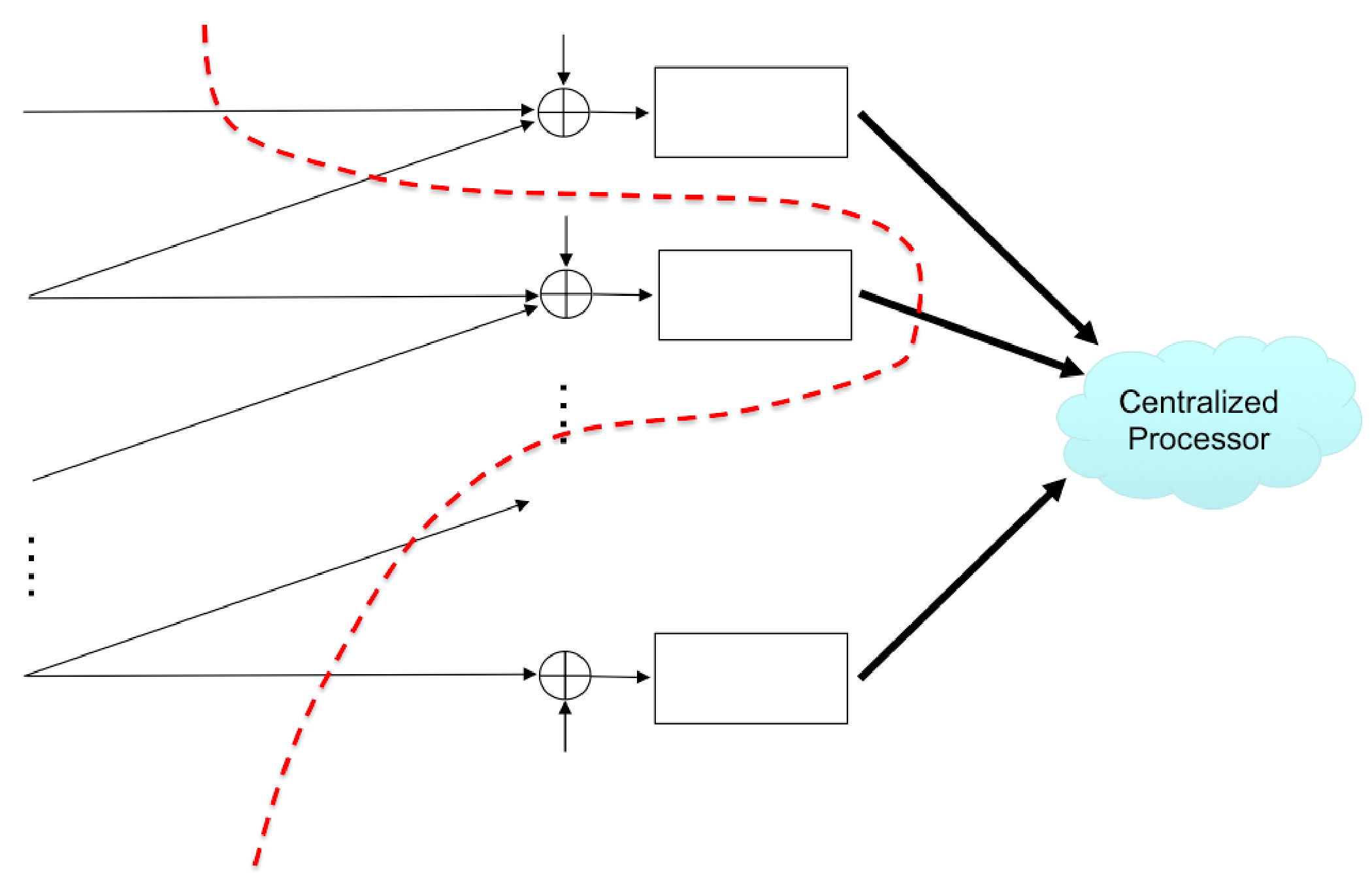}
\centering
\put(50,55){\small $Y_1:\hat{Y}_1$}
\put(50,42.5){\small $Y_2:\hat{Y}_2$}
\put(49.3,14.5){\small $Y_L:\hat{Y}_L$}
\put(-4,56){\small $X_1$}
\put(-4,42){\small $X_2$}
\put(-4,14){\small $X_L$}
\put(36,62){\small $Z_1$}
\put(36,47){\small $Z_2$}
\put(35,9){\small $Z_L$}
\put(10,58){\small $h_{11}$}
\put(11.5,49){\small $h_{21}$}
\put(17.5,38){\small $h_{32}$}
\put(25,44.5){\small $h_{22}$}
\put(25,13){\small $h_{LL}$}
\put(69,52){\small $C_1$}
\put(68,43){\small $C_2$}
\put(68,18){\small $C_L$}
\put(80,41){\small $\hat{Y}_1$}
\put(73,34){\small $\hat{Y}_2$}
\put(77,25){\small $\hat{Y}_L$}
\put(13, 2){$\mathcal{S}$}
\put(22, 2){$\mathcal{S}^c$}
\end{overpic}
\caption{Choice of cut-set for the Wyner model. The cut goes through 
the access link or the backhaul link depending on whether
$\frac{1}{2} \log(\mathsf{SNR}_i)$ is greater or smaller than $C_i$.}
\label{cutset}
\end{figure}

\begin{IEEEproof}
To show that the difference between the achievable sum rate in
(\ref{sum_rate_asymmetric}) and the sum capacity is bounded by 
a constant gap, we first write down the cut-set bound 
\cite[Theorem~14.10.1]{EIT:Cover} for the Wyner channel model. 
Let $\mathcal{L} = \{1, 2, \cdots, L\}$
represent the index set for the base-stations. We partition
$\mathcal{L}$ into two sets, $\mathcal{S}$ and $\mathcal{S}^c$, 
and only consider cuts for which all the $X_i$'s 
and a selected subset of $Y_i$'s, with $i \in \mathcal{S}$, are on one
side, and the rest of the $Y_i$'s, with $i \in \mathcal{S}^c$, and the
central processor are on the other side. 
Denote $\mathbf{X}(\mathcal{L}) = \{X_i \ | \ i \in \mathcal{L} \}$
and $\mathbf{Y}(\mathcal{S}^c) = \{Y_i \ | \ i \in \mathcal{S}^c \}$.
We have an upper bound to the cut-set bound as follows:
\begin{eqnarray} 
R_{\mathrm{sum}}^{\mathrm{cut-set}} & \le & 
\max_{p(\mathbf{X}(\mathcal{L}))} \min_{\mathcal{S} \subseteq \mathcal{L}}
\left \{ I\left(\mathbf{X}(\mathcal{L}); \mathbf{Y}(\mathcal{S}^c)
\right) + \sum_{i \in \mathcal{S}}C_i \right \} \nonumber \\
& \le & \min_{\mathcal{S} \subseteq \mathcal{L}} \left \{
\max_{p(\mathbf{X}(\mathcal{L}))} I\left(\mathbf{X}(\mathcal{L}); \mathbf{Y}(\mathcal{S}^c) \right) + \sum_{i \in \mathcal{S}}C_i \right \}, \nonumber \\ \label{cutset_bound}
\end{eqnarray} where the maximization is taken over all the possible
joint distributions of $\mathbf{X}(\mathcal{L})$, and the
minimization is taken over $2^L$ different choices of cuts that
separate the $L$ transmitters and the centralized processor. 

Evaluate the mutual-information term in (\ref{cutset_bound}) as follows: 
\begin{eqnarray}
\lefteqn{I\left(\mathbf{X}(\mathcal{L}); \mathbf{Y}(\mathcal{S}^c) \right) } \nonumber \\
&=& h \left(\mathbf{Y}(\mathcal{S}^c)\right) - \sum_{i \in \mathcal{S}^c} h(Z_i) \nonumber \\
&\le& \sum_{i \in \mathcal{S}^c} \left(h(Y_i) - h(Z_i) \right) \nonumber \\
&\stackrel{(a)}{\le}& \sum_{i \in \mathcal{S}^c} \frac{1}{2} \log (1 + \mathsf{SNR}_i + \mathsf{INR}_{i+1, i}) \nonumber \\
&\stackrel{(b)}{\le}& \left\{
\begin{array}{ll}
\sum_{i \in \mathcal{S}^c} \frac{1}{2} \log (1 + \mathsf{SNR}_i) +
\frac{|\mathcal{S}^c|}{2}, & {\rm if\ } L \in S \\
\sum_{i \in \mathcal{S}^c} \frac{1}{2} \log (1 + \mathsf{SNR}_i) +
\frac{|\mathcal{S}^c|-1}{2}, & {\rm if\ } L \in S^c \\
\end{array}
\right.
\end{eqnarray}
where (a) holds because Gaussian distribution maximizes the entropy $h(Y_i)$ and (b) is due to the fact that $\mathsf{INR}_{i+1, i} \le \mathsf{SNR}_i$ thus $\frac{1}{2} \log (1 + \mathsf{SNR}_i + \mathsf{INR}_{i+1, i}) \le \frac{1}{2} \log (1 + \mathsf{SNR}_i) + \frac{1}{2}$
for $i=1,\cdots,L-1$. For $i=L$, $h(Y_L)-h(Z_L)=\frac{1}{2} \log(1+\mathsf{SNR}_L)$.


Now, we are going to choose a particular cut to further upper bound
the cut-set bound. We compare $\frac{1}{2} \log(\mathsf{SNR}_i)$
at each base-station with the backhaul capacity $C_i$, and let the cut
go through either the user-base-station link or the backhaul link,
whichever corresponds to the smaller of the two quantities above, as
shown in Fig.~\ref{cutset}. 
More precisely, define 
\begin{equation} \label{partition}
\mathcal{S} = \left\{ i \ \left| \frac{1}{2} \log \mathsf{SNR}_i \ge C_i \right.
\right\}.
\end{equation}
With this particular cut, we can now bound the difference between the cut-set
bound and the achievable sum rate. 
%
%
First, consider the case that $L \in S^c$, we use the achievable sum rate
as given in (\ref{sum_rate_asymmetric}):
\begin{eqnarray}
\lefteqn{R_{\mathrm{sum}}^{\mathrm{cut-set}} - R_{\mathrm{sum}}} \nonumber \\
&\le& \sum_{i \in \mathcal{S}^c} \left\{ \frac{1}{2} \log (1 + \mathsf{SNR}_i) -  \frac{1}{2} \log \left(  \frac{1 + \mathsf{SNR}_i}{1 + 2^{-2C_i}\mathsf{SNR}_i}\right) \right\} \nonumber \\
&& + \sum_{i \in \mathcal{S}} \left \{C_i -   \frac{1}{2} \log \left(
\frac{1 + \mathsf{SNR}_i}{1 + 2^{-2C_i}\mathsf{SNR}_i}\right) \right\}
+ \frac{|\mathcal{S}^c|-1}{2} \nonumber \\
&=& \sum_{i \in \mathcal{S}^c} \frac{1}{2} \log \left(1 + 2^{-2C_i}\mathsf{SNR}_i \right)  \nonumber \\
&&+ \sum_{i \in \mathcal{S}} \frac{1}{2} \log \left(\frac{2^{2C_i} +
\mathsf{SNR}_i}{1 + \mathsf{SNR}_i} \right) + \frac{|\mathcal{S}^c|-1}{2}  \nonumber \\
&\stackrel{(a)}{\le}& \frac{|\mathcal{S}^c|}{2} +
\frac{|\mathcal{S}|}{2} +
\frac{|\mathcal{S}^c|-1}{2}  
\nonumber \\
&\le& L - \frac{1}{2} \label{L-0.5_1}
\end{eqnarray}
where in (a) we used the definition of the cut (\ref{partition}),
i.e., $\mathsf{SNR}_i \le 2^{2C_i}$ for $i \in \mathcal{S}^c$, 
and $\mathsf{SNR}_i \ge 2^{2C_i}$ for $i \in \mathcal{S}$, 
and the last inequality is due to the fact that $|\mathcal{S}^c| + |\mathcal{S}| = L$
and $|\mathcal{S}^c| \le L$. 

Now, consider the case that $L \in \mathcal{S}$. We tighten the sum
rate expression (\ref{sum_rate_asymmetric}) by noticing that
since by the definition of $\mathcal{S}$ we have $\frac{1}{2} \log \mathsf{SNR}_L 
\ge C_L$, so by (\ref{RL_improved}) we have $R_L=C_L$. In this case,
\begin{eqnarray}
\lefteqn{R_{\mathrm{sum}}^{\mathrm{cut-set}} - R_{\mathrm{sum}}} \nonumber \\
& \le & \sum_{i \in \mathcal{S}^c} \left\{ \frac{1}{2} \log (1 + \mathsf{SNR}_i) -  \frac{1}{2} \log \left(  \frac{1 + \mathsf{SNR}_i}{1 + 2^{-2C_i}\mathsf{SNR}_i}\right) \right\} \nonumber \\
&& + \sum_{i \in \mathcal{S}\setminus\{L\}} \left \{C_i -   \frac{1}{2} \log \left(
\frac{1 + \mathsf{SNR}_i}{1 + 2^{-2C_i}\mathsf{SNR}_i}\right) \right\}
+ \frac{|\mathcal{S}^c|}{2} \nonumber \\
& \le & \frac{|\mathcal{S}^c|}{2} + \frac{|\mathcal{S}|-1}{2} + \frac{|\mathcal{S}^c|}{2} \le  L - \frac{1}{2} \label{L-0.5_2}
\end{eqnarray}
This completes the proof.  \end{IEEEproof}
\vspace{1em}

It turns out that even without Wyner-Ziv compress-and-forward,
it is still possible to achieve the sum capacity
to within a constant, albeit larger, gap under the same
weak-interference condition.

\begin{theo}\label{theo:asymmetric_no_wynerziv}
For a multicell processing Wyner model shown in
Fig.~\ref{asymmetric}, in the weak-interference regime of
$\mathsf{INR}_{i+1, i} \le \mathsf{SNR}_i, i=1, 2, \cdots, L-1$, the
per-base-station SIC scheme without Wyner-Ziv coding achieves a sum rate
that is within 
$\frac{1}{2}(1+\log(3))L-\frac{1}{2}$ 
bits of the sum capacity.
\end{theo}


\begin{IEEEproof}
Applying Theorem~\ref{theo: per-base-station-SIC-NOWZ} to the Wyner
model of Fig.~\ref{asymmetric} with the decoding order from user $L$
to user $1$, it is easy to see that the achievable rate of user $L$
remains the same 
while the rates of other users become 
\begin{equation}
\label{no_wz_rate}
R_i = \frac{1}{2} \log \frac{1 + 2^{-2C_i}\mathsf{INR}_{i+1, i}  + \mathsf{SNR}_i}{1 + 2^{-2C_i}\mathsf{INR}_{i+1, i}  + 2^{-2C_i}\mathsf{SNR}_i}.
\end{equation}
Under the weak-interference condition $\mathsf{INR}_{i+1, i}
\le \mathsf{SNR}_i$, when $i \in {\cal S}^c$, we have
$0 \le 2^{-2C_i}\mathsf{INR}_{i+1, i} \le 1$, and when  
when $i \in {\cal S}$, we have
$0 \le 2^{-2C_i}\mathsf{INR}_{i+1, i} \le 2^{-2C_i}\mathsf{SNR}_{i}$.
Replace the $R_i$ expressions in (\ref{L-0.5_1}) and (\ref{L-0.5_2})
by (\ref{no_wz_rate}) while noting the inequalities above, we obtain the
$\frac{1}{2}(1+\log(3))L-\frac{1}{2}$ bound.
\end{IEEEproof}

Note that a looser bound can be obtained by noting that 
under the weak-interference condition $\mathsf{INR}_{i+1, i}
\le \mathsf{SNR}_i$, the achievable rate without Wyner-Ziv coding
(\ref{no_wz_rate}) is already within $\frac{1}{2}$ bits of the 
rate with Wyner-Ziv coding as in (\ref{rate_asymmetric_Gaussian}).
As a result, a looser sum-rate gap to the cut-set upper bound 
can immediately be obtained as $L-\frac{1}{2} + \frac{L-1}{2} =
\frac{3}{2}L-1$ bits\footnote{The authors wish to thank Yuhan Zhou to
point out the tighter bound in Theorem \ref{theo:asymmetric_no_wynerziv}.
}.


\section{Optimal Rate Allocation with a Total Backhaul Capacity Constraint}

A practical system may have a constraint on the sum capacity of all digital
backhaul links, for example, when the backhaul links are implemented in a
wireless medium with shared bandwidth. So, it may be of interest to optimize
the allocation of backhaul capabilities among the base-stations in order to
achieve an overall maximum sum rate under a total backhaul capacity constraint.
The structure of the solution of such an optimization can also yield useful 
insight. The optimization problem can be formulated as the following:
\begin{eqnarray}
\label{opt_backhaul}
& \mathrm{maximize}  \quad & \sum_{k=1}^{L}R_k   \\
& \mathrm{subject \; to} \quad & C_k \ge 0, \;\; k=1, 2, \cdots, L \nonumber \\
&& \sum_{k=1}^{L}C_k  \le C  \nonumber
\end{eqnarray}
where $R_k, k=1,2,\cdots, L$ are functions of $C_k$ as derived in
Theorem~\ref{rate:per_BS_SIC_WZ}, and $C > 0 $ is the total available
backhaul capacity. The following theorem provides an optimal solution
to the above optimization problem for the per-base-station SIC scheme
with Wyner-Ziv coding.

\begin{theo} \label{rate_allocation_theo}
For the uplink multicell joint processing model shown in
Fig.~\ref{cloud_model}, with the per-base-station SIC at the central
processor with Wyner-Ziv coding, the optimal allocation of backhaul
link capacities subject to a total backhaul capacity constraint $C$
is given by \begin{equation}
\label{optimal_allocation}
C_k^* = \max \left\{ \frac{1}{2}\log(\mathsf{\overline{SINR}}_k) - \alpha , 0 \right\},
\end{equation}
where $\alpha$ is chosen such that $\sum_{k=1}^{L} C_k^* = C$.
\end{theo}
\begin{IEEEproof}
Substituting the rate expression (\ref{rate: improved-per-base-station-SIC}) for $R_k$
into the optimization problem (\ref{opt_backhaul}),
we obtain the following equivalent minimization problem:
\begin{eqnarray}
\label{opt_prob}
& \mathrm{minimize}  \quad & \sum_{k=1}^{L}\frac{1}{2} \log \left(
1 + 2^{-2C_k} \mathsf{\overline{SINR}}_k \right)   \\
& \mathrm{subject \; to} \quad & C_k \ge 0, \;\; k=1, 2, \cdots, L. \nonumber \\
&& \sum_{k=1}^{L}C_k  \le C  \nonumber
\end{eqnarray}
It can be easily seen that (\ref{opt_prob}) is a convex optimization problem,
as the constraints are linear and the objective function is the sum of
convex functions, as can be verified by taking their second derivatives.


Now introducing Lagrange multipliers {\boldmath $\nu$} $\in
\mathbb{R}_+^L$ for the positivity constraints $C_k \ge 0, k=1, 2,
\cdots, L$, and $\lambda \in \mathbb{R}_+$ for the
backhaul sum-capacity constraint $\sum_{k=1}^{L} C_k \le C$, we
form the Lagrangian
\begin{multline}
L(C_k,{\boldmath \boldsymbol{\nu}},\lambda) =
	\sum_{k=1}^{L}\frac{1}{2} \log \left( 1 + 2^{-2C_k}
	\mathsf{\overline{SINR}}_k \right)  \\
- \sum_{k=1}^L \nu_k C_k + \lambda \left( \sum_{k=1}^L C_k - C \right)
\end{multline}
Taking the derivative of the above with respect to $C_k$, we
obtain the following Karush-Kuhn-Tucker (KKT) condition
\begin{equation}
- \frac{ 2^{-2C_k^*}\mathsf{\overline{SINR}}_k}
	{1 + 2^{-2C_k^*}\mathsf{\overline{SINR}}_k} - \nu_k + \lambda = 0,
\label{KKT}
\end{equation}
for the optimal $C_k^*$, where $k=1, 2, \cdots, L$. Note that $\nu_k=0$
whenever $C_k > 0$. Now, the optimal $C_k^*$ must satisfy
the backhaul sum-capacity constraint $\sum_{k=1}^{L}C_k^* \le C$
with equality, because the objective of the minimization $R_k$
monotonically increases with $C_k$. Solving the condition (\ref{KKT})
together with the fact that $\sum_{k=1}^{L}C_k^* = C$ gives the
following optimal $C_k^*$:
\begin{equation}
C_k^* = \max \left\{ \frac{1}{2}\log\frac{\mathsf{\overline{SINR}}_k}{\beta}, 0 \right\},
\label{optimal_allocation_2}
\end{equation}
where $\beta$ is chosen such that $\sum_{k=1}^{L} C_k^* = C $.
This is equivalent to (\ref{optimal_allocation}).
\end{IEEEproof}

\vspace{0.1cm}

An interpretation of (\ref{optimal_allocation_2}) is that whenever
the SINR of user $k$ is above a threshold $\beta$, $\frac{1}{2}
\log\frac{\mathsf{\overline{SINR}}_k}{\beta}$
bits of the backhaul link should be allocated to user $k$. Otherwise,
this user should not be active as far as maximizing the uplink sum 
rate is concerned.
This optimal rate allocation is quite similar to the classic
water-filling solution for the sum-capacity maximization problem for 
a parallel set of Gaussian channels, in which more power (backhaul
capacity in this case) is assigned to users with a better channel.

When written as (\ref{optimal_allocation}), the optimal backhaul
capacity allocation can be interpreted as follows:
$C_k = \frac{1}{2}\log(\mathsf{\overline{SINR}}_k)$ can be thought of as
the nominal optimal backhaul link capacity. If the total backhaul
capacity is above (or below) the nominal $\sum_k
\frac{1}{2}\log(\mathsf{\overline{SINR}}_k)$, then the extra capacity must be
distributed (or taken away) from each base-station equally. In other
words, all base-station should nominally operate at the point 1/2 bits
away from the SIC limit (as shown in Fig.~\ref{Rk_vs_Ck}). If more
(or less) backhaul capacity is available than the nominal value,
all base-stations
should move above (or below) that operating point in the same manner.

Finally, we remark that the decoding order at the centralized
processor plays an important role in the optimal rate allocation.
Different decoding orders result in different rate expressions in
Theorem~\ref{theo: per-base-station-SIC-WZ} and thus different rate
allocations in Theorem~\ref{rate_allocation_theo}, and as a
consequence different achievable sum rates. In order to determine the
decoding order that results in the largest sum rate (or the
maximum network utility), we need to exhaustively search over $K!$
different decoding orders. This is a nontrivial
problem that is also encountered in other contexts involving
successive decoding such as in V-BLAST \cite{vblast_decoding_order_1}.

\section{Numerical Examples}

\subsection{Two-User Symmetric Scenario}


\begin{figure*}[t]
\centering

\subfigure[$\mathsf{SNR}=30$dB, $\mathsf{INR}=20$dB, $C_1 = C_2 =5$ bits]{
\includegraphics[width=3.1in]{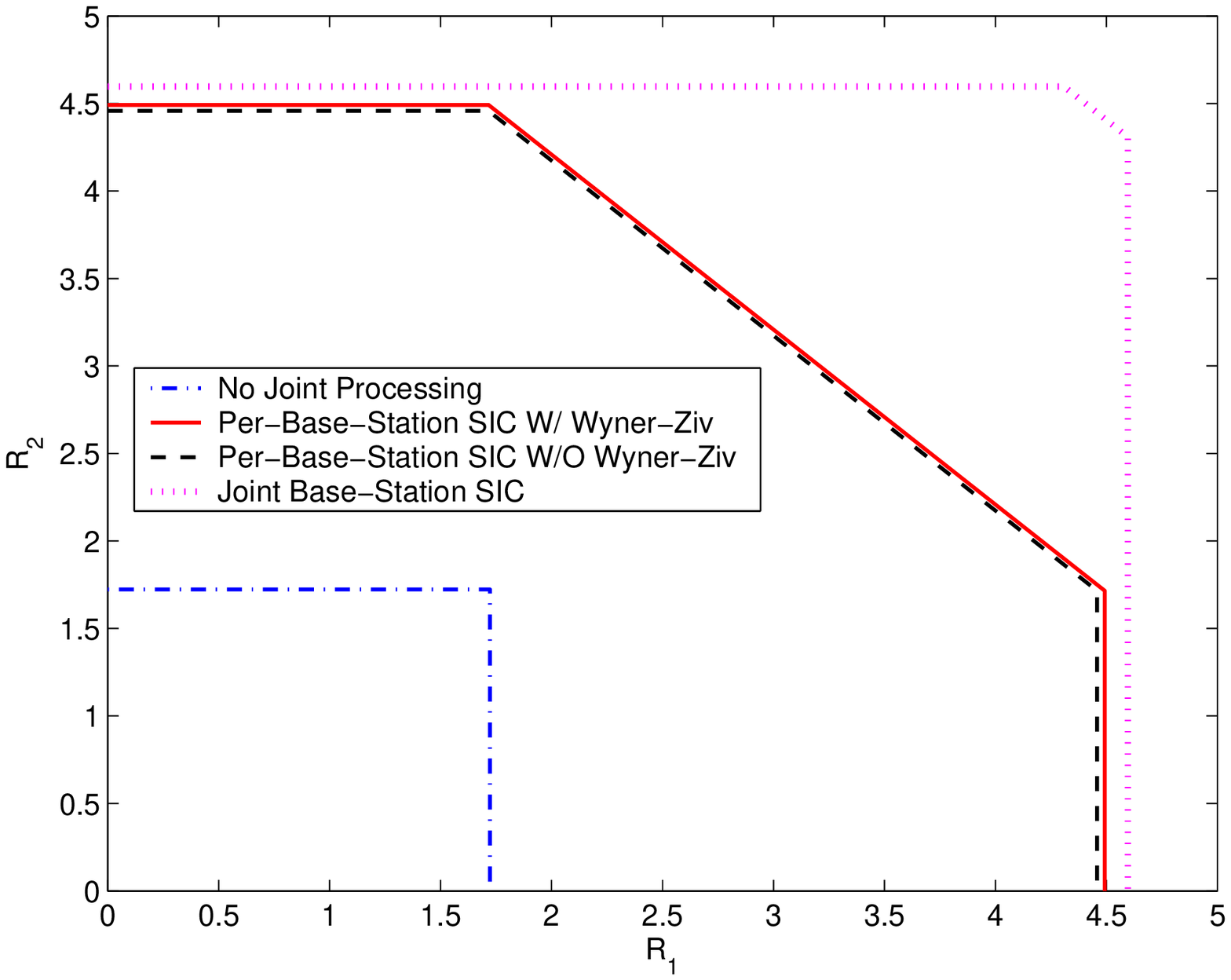}
\label{30-20-5bit} } 
\subfigure[$\mathsf{SNR}=30$dB, $\mathsf{INR}=5$dB, $C =5$ bits]{
\includegraphics[width=3.1in]{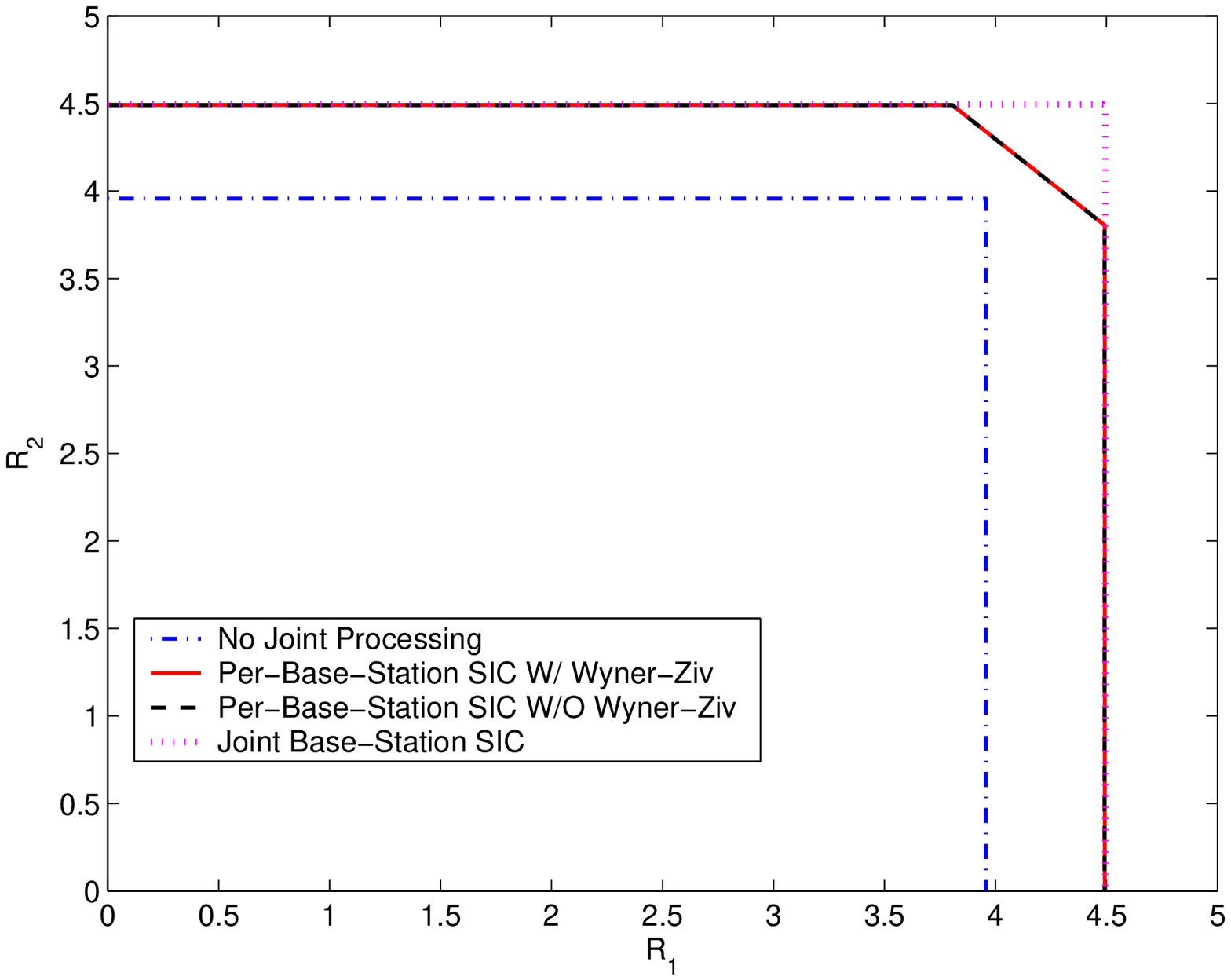}
\label{30-5-5bit} } 

\vspace{1.5em}

\subfigure[$\mathsf{SNR}=30$dB, $\mathsf{INR}=20$dB, $C=10$ bits]{
\includegraphics[width=3.1in]{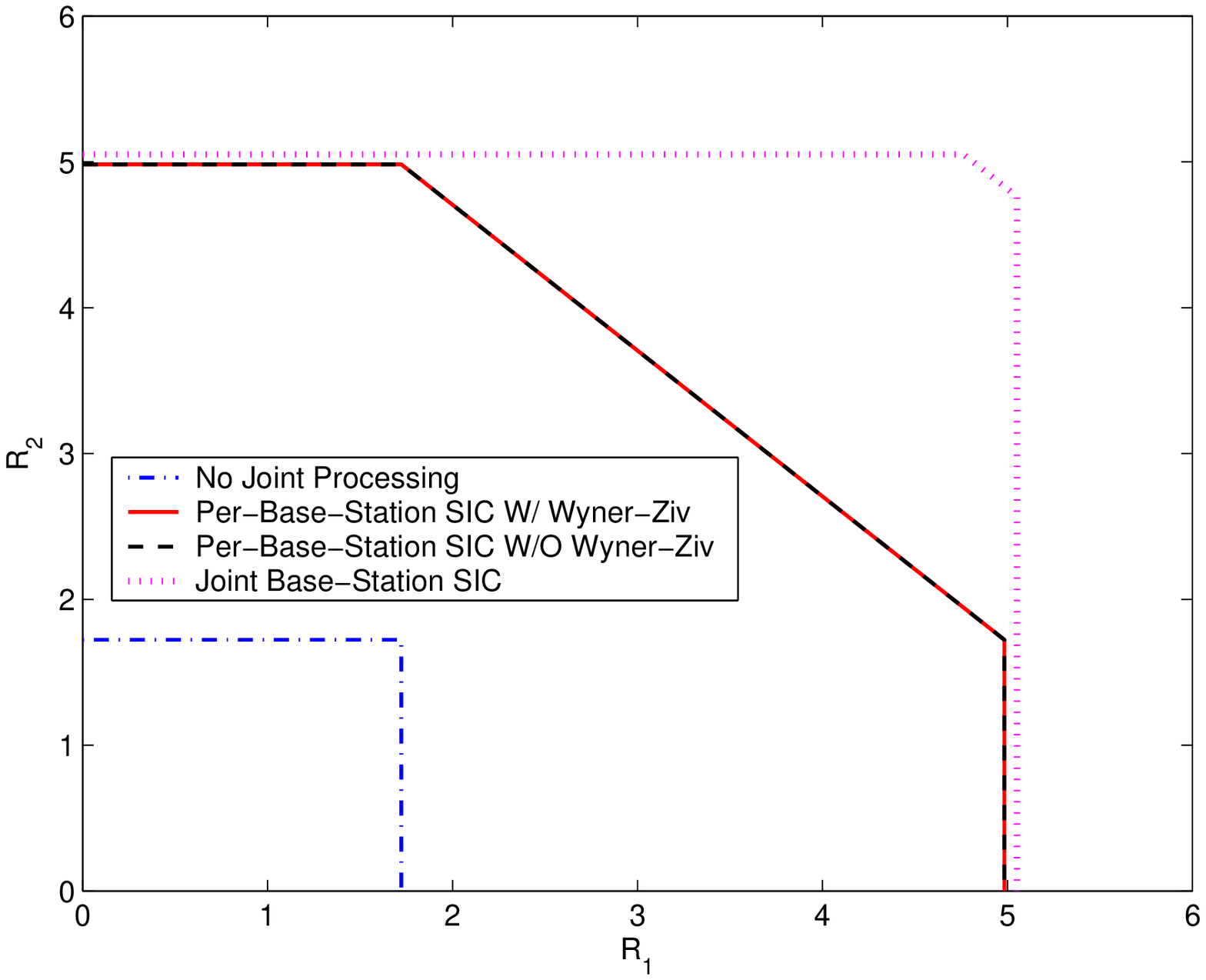}
\label{30-20-10bit} } 
\subfigure[$\mathsf{SNR}=30$dB, $\mathsf{INR}=20$dB, $C =2$ bits]{
\includegraphics[width=3.1in]{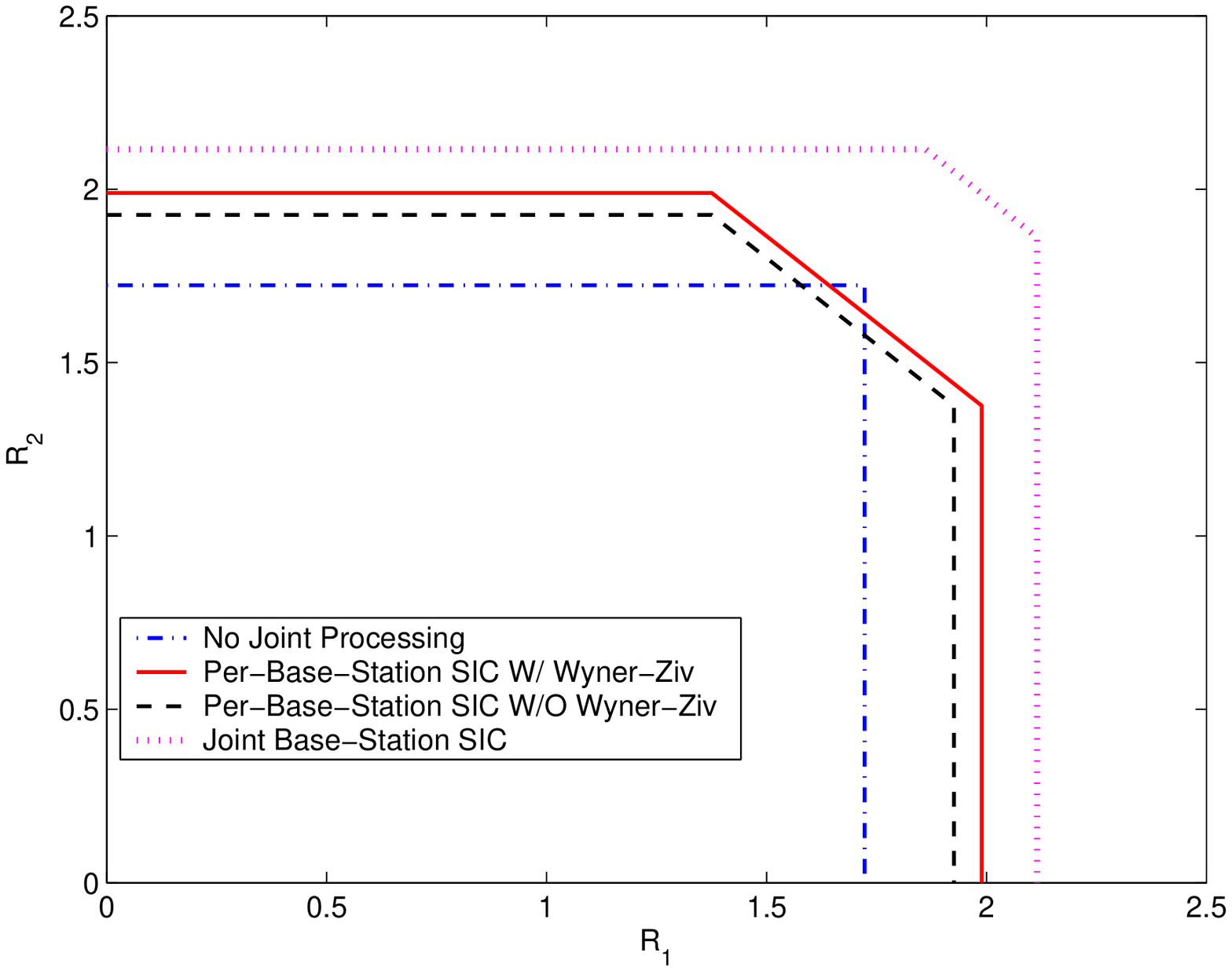}
\label{30-20-2bit} }
\caption{Comparison of the proposed achievability scheme and another two schemes} \label{fig:common_msg_intersection}
\end{figure*}

To obtain numerical insight on the per-base-station SIC-based schemes
proposed in this paper, the achievable rate regions of
Theorem~\ref{theo: per-base-station-SIC-WZ} and Theorem~\ref{theo:
per-base-station-SIC-NOWZ}
are compared with that obtained by two other schemes: 1)
single-user decoding without joint processing; 2) joint base-station
processing as defined in (\ref{rate: joint-BS SIC}) in Section
\ref{joint-bs-processing}, for a two-user symmetric scenario where
$L=2$, $P_1 = P_2 = N_0 =1$, $\mathsf{SNR}=h_{11}^{2}=h_{22}^2$,
$\mathsf{INR}=h_{12}^2 = h_{21}^2$, and $C=C_1= C_2$.  Under the
symmetric setting, both Theorem~\ref{theo: per-base-station-SIC-WZ}
and Theorem~\ref{theo: per-base-station-SIC-NOWZ} give two
symmetric achievable rate pairs depending on the decoding order.
Time-sharing of the two achievable rate pairs gives a pentagon shaped
achievable rate region.

Single-user decoding without joint processing is considered as a
baseline, in which each receiver
decodes its own signal while treating the other user's signal as
noise.  This gives the following achievable rate pair
\begin{equation}
 R_1  = R_2 = \min \left\{ \D \frac{1}{2} \log \D \left( 1 + \frac{\mathsf{SNR}}{1 +
		\mathsf{INR}} \right), C \right\}
\end{equation}
which in the symmetric setting results in a square shaped achievable
rate region with ($R_1, R_2$) as the top-right corner.

We also compare with the joint-base-station processing SIC scheme of
(\ref{rate: joint-BS SIC}). We restrict ourselves to
symmetric quantization noise levels here.  The quantization noise level
$\mathsf{q}$ is chosen such that the resulting average backhaul link
capacity is $C$, i.e., 
\begin{equation}
I(Y_1; \hat{Y}_1) + I(Y_2; \hat{Y}_1|\hat{Y}_1) = 2C.
\end{equation}
This condition gives the following analytic expression for the 
quantization noise level:
\begin{equation} \label{q_level_joint_bs_processing}
\mathsf{q} = \frac{a+\sqrt{4b+2^{4C} \left(a^2-4b\right)}}{2^{4C}-1},
\end{equation}
where $a = 1 + \mathsf{SNR} + \mathsf{INR}$,
and $b = \mathsf{SNR} \cdot \mathsf{INR}$.
Plots of the joint-base-station SIC are then obtained using
(\ref{rate: joint-BS SIC}) with this quantization noise level. 


The achievable rate regions obtained above are compared for the
following channel settings:
\begin{itemize}
\item $\mathsf{SNR}=30$dB, $\mathsf{INR}=20$dB, $C =5$ bits;
\item $\mathsf{SNR}=30$dB, $\mathsf{INR}=5$dB, $C =5$ bits;
\item $\mathsf{SNR}=30$dB, $\mathsf{INR}=20$dB, $C =10$ bits;
\item $\mathsf{SNR}=30$dB, $\mathsf{INR}=20$dB, $C =2$ bits.
\end{itemize}
Fig.~\ref{30-20-5bit} shows 
that at relatively strong interference level $\mathsf{INR}=20$dB,
the proposed per-base-station SIC schemes (both with and without
Wyner-Ziv compression) expand the baseline achievable rate region by
about $2.8$ bits on either of the individual rates and on the sum rate. The
joint base-station SIC regions further outperforms the proposed
scheme in sum rate by about $2.5$ bits due to the benefits of joint
decoding.

However, when the interfering links are weak, as shown in
Fig.~\ref{30-5-5bit} where $\mathsf{INR}=5$dB, all four
achievable rate regions are close to each other. This is the regime
where treating interference as noise is close to optimal, so multicell
processing does not provide significant benefits.

In the above two examples, the capacities of the backhaul links are
already quite abundant, since they are set to be the rate supported
by the direct links: $\frac{1}{2}\log(1 + \mathsf{SNR}) \approx 5$ bits. In
Fig.~\ref{30-20-10bit}, we further increase the backhaul capacity to
10 bits, and show that doing so does not significantly improve the
achievable rate region for either proposed SIC-based schemes or the joint base-station processing scheme.


Lastly, we decrease the backhaul capacity from 5 bits to 2 bits.
Interestingly, this is a situation in which the base-line scheme can
outperform per-base-station SIC as shown in Fig.~\ref{30-20-2bit}. 
Therefore, the proposed per-base-station schemes can be inefficient 
in term of sum rate, when the backhaul rates are very limited.
Observe that the largest sum rate is still obtained with joint base-station
processing.  

%
%
%
%
%

\subsection{Multicell OFDMA Network}

\begin{figure}[t]
\centering
\includegraphics[width=3.1in]{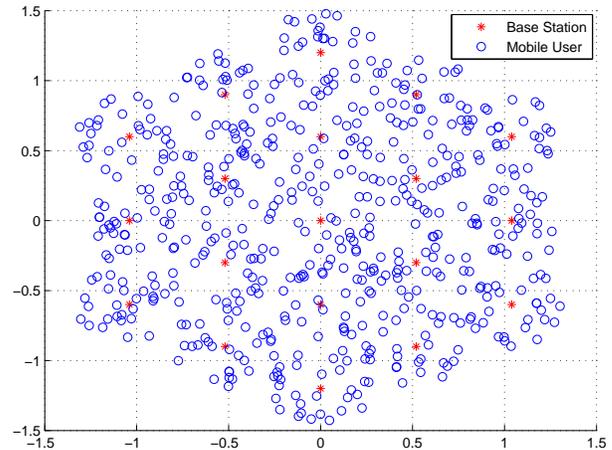}
\caption{A cellular topology with $19$ cells, $3$ sectors per cell,
and $10$ users per sector placed randomly. The distance between the 
neighboring base-stations is set to $600$m.}
\label{19cell}
\end{figure}

To further understand the performance of the proposed per-base-station
SIC scheme in practical systems, in this section, the achievable rates
of the two variations of the per-base-station SIC, i.e., with and
without Wyner-Ziv coding, 
are evaluated for a wireless cellular network setup with $19$ cells, 
$3$ sectors per cell, and $10$ users per sector,  where an
orthogonal frequency-division multiple-access (OFDMA) scheme with $64$
tones over a fixed $10$Mbps bandwidth is employed. The cellular
topology is shown in Fig.~\ref{19cell}.  A 19-cell wrap-around
layout is used to ensure uniform interference statistics throughout
the network. 
The assignments of frequency tones
to users within each cell are non-overlapping. As a result, users
experience only intercell interference and no intracell interference.
Both the base-stations and the mobile users are equipped with a single
antenna each. Each of the $19$ base-stations is
connected to the centralized processor via a rate-limited  backhaul
link. Perfect channel estimation is assumed, and the CSI is made
available to all base-stations and to the centralized processor. In
the simulation, uniform power allocation of $-27$dBm/Hz is assumed at
all the mobile users. For convenience, a round-robin scheduler is used for user
assignment. The base-station-to-base-station distance is set to $600$m
corresponding to a typical urban deployment. Detailed system
parameters are outlined in Table~\ref{table_parameter}.

\begin{table}[t]
  \caption{Wireless Cellular Model Parameters}
  \centering
  \begin{tabular}{@{} |c|c| @{}}
    \hline
    Cellular Layout & Hexagonal, 19 cells, 3 sectors/cell \\
    \hline
    BS-to-BS Distance & 600 m \\     \hline
    Frequency Reuse & 1 \\     \hline
    Number of User per Sector & 10 \\     \hline
    Channel Bandwidth & 10 MHz \\     \hline
    MS Transmit PSD & $-27$ dBm/Hz \\     \hline
    Antenna Gain & 15 dBi \\     \hline
    Background Noise & $-169$ dBm/Hz \\     \hline
    Noise Figure & $7$ dB \\     \hline
    Multipath Time Delay Profile & ITU-R M.1225 PedA \\     \hline
    Distance-dependent Path Loss & $128.1 + 37.6\log_{10}(d)$ \\     \hline
    Number of Tones & 64 \\     \hline
  \end{tabular}
  \label{table_parameter}
\end{table}

In the first part of the 
simulation, the capacities of the backhaul links are fixed per
base-station and uniformly distributed across the frequency tones, 
e.g., if the capacity of a backhaul link is $64$Mbps, each frequency 
tone is assumed to have a backhaul of $64\mathrm{Mbps}/64 = 1\mathrm{Mbps}$. 
Cumulative distribution function (CDF) of the user rates is plotted
in order to visualize the performance gain of the proposed schemes 
over a baseline system, in which base-stations decode the user
messages without joint processing at the centralized processor. For
the proposed per-base-station SIC scheme, to account for the fairness
among users, the decoding order across the cells is chosen to be in
decreasing order of the user SINRs (prior to SIC). This decoding order
is adopted on each of the OFDM tones independently.

\begin{figure} [t]
\centering
\includegraphics[width=3.4in]{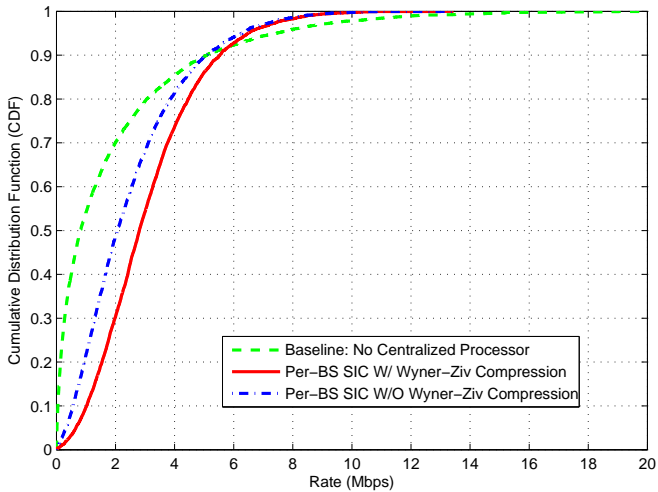}
\caption{CDF of user rates with 180Mbps backhaul per base-station uniformly allocated
across the frequencies} \label{180Mbps}
\end{figure}

Fig.~\ref{180Mbps} shows the CDF plots of user rates with a backhaul
capacity at each base-station of $180$Mbps (i.e., $60$Mbps per
sector). It is seen that the two per-base-station SIC schemes (with or
without Wyner-Ziv coding) both significantly outperform
the baseline system. The user rate at $50$th-percentile is
around $0.8$Mbps for the baseline, $2.1$Mbps for the per-base-station
SIC scheme without Wyner-Ziv coding, and $2.8$Mbps for the
per-base-station SIC scheme with Wyner-Ziv coding. Compared with
the baseline curve, the per-base-station SIC curves also have a better
distribution of user rates in terms of fairness.
There is a noticeable performance gap between the SIC curve with
Wyner-Ziv compression and without Wyner-Ziv-compression. This gap is
due to the compression gain brought by side information.

\begin{figure} [t]
\centering
\includegraphics[width=3.4in]{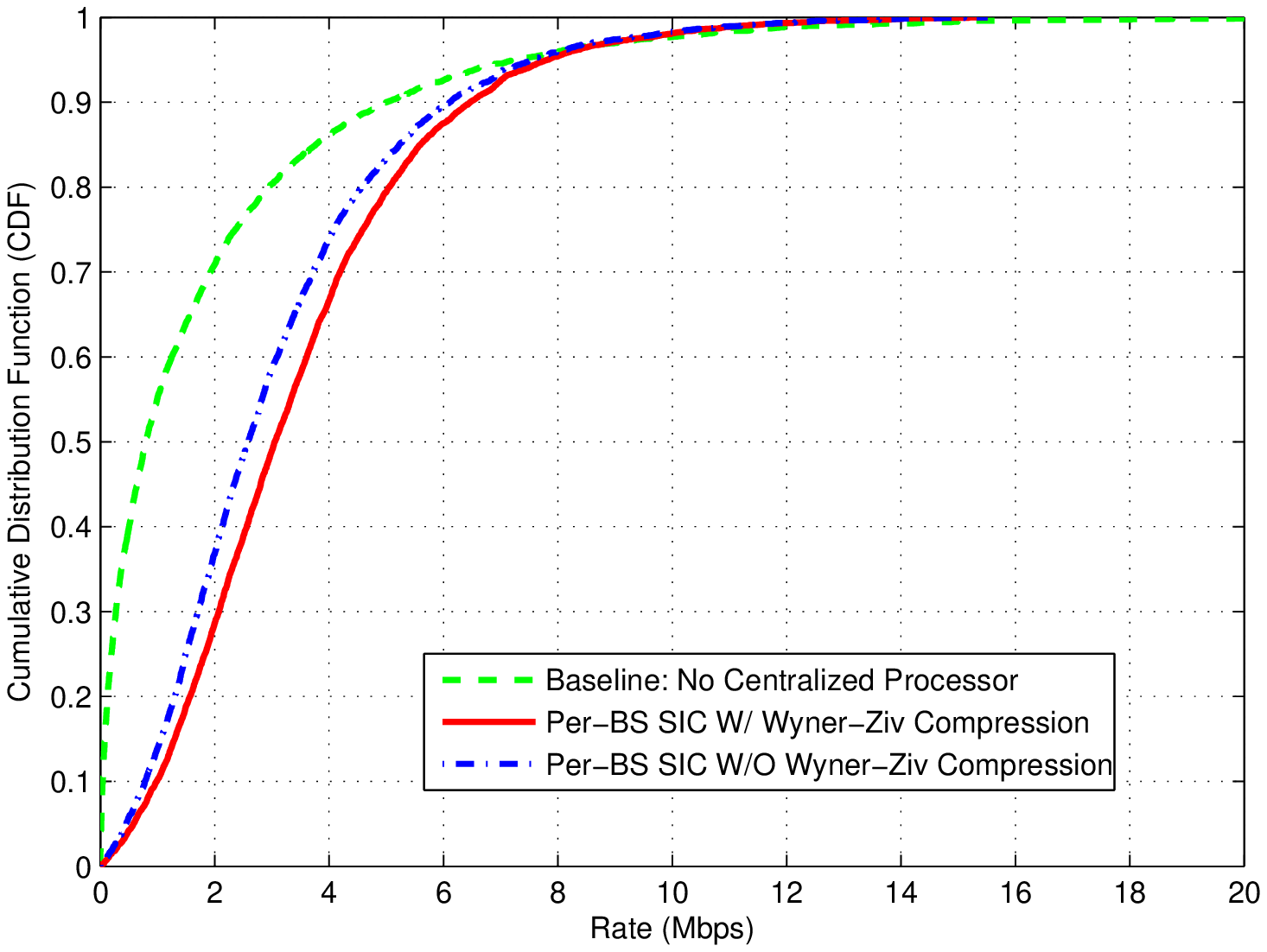}
\caption{CDF of user rates with 270Mbps backhaul per base-station uniformly allocated
across the frequencies} \label{270Mbps}
\end{figure}

\begin{figure} [t]
\centering
\includegraphics[width=3.4in]{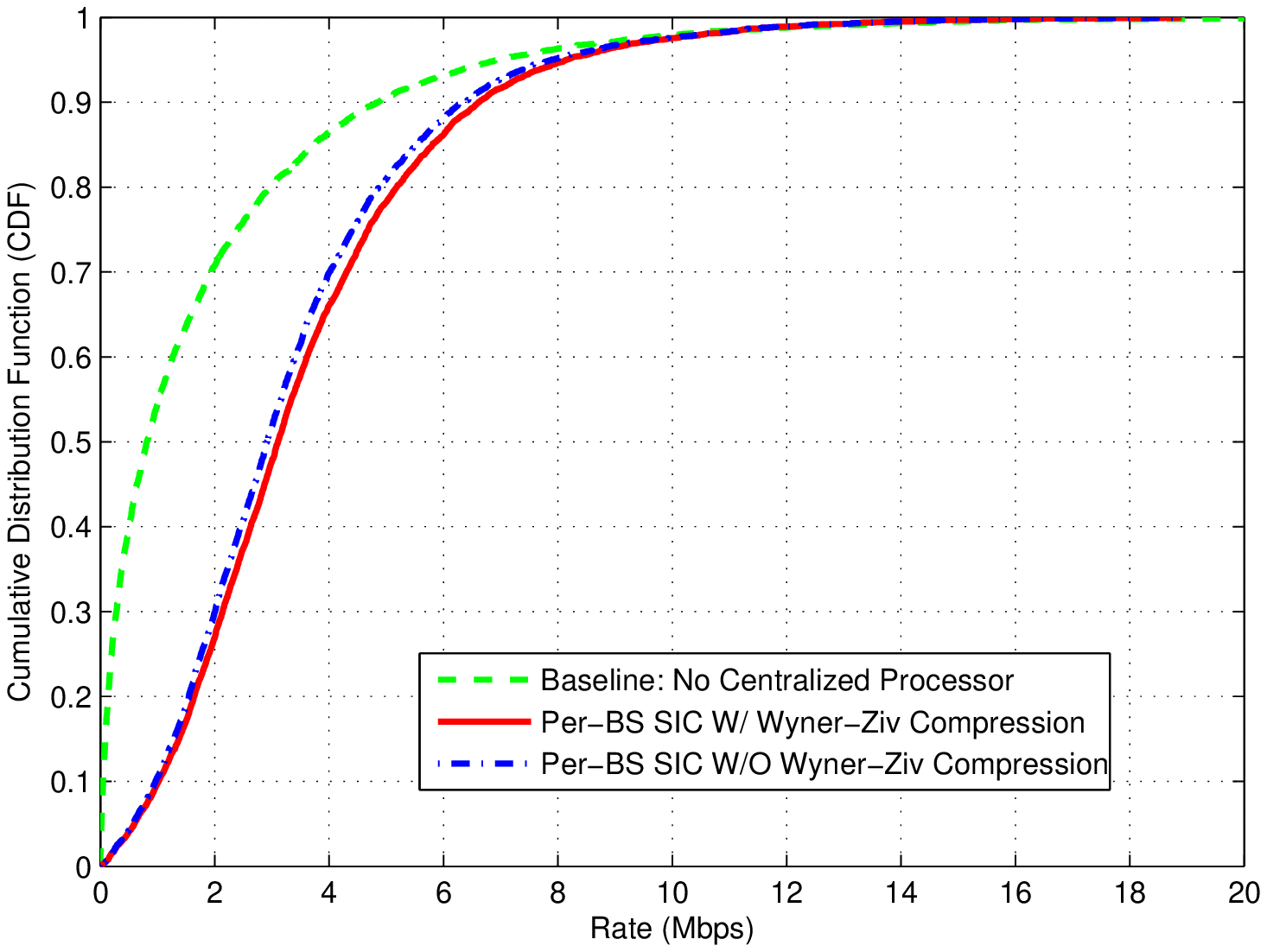}
\caption{CDF of user rates with 360Mbps backhaul per base-station uniformly allocated
across the frequencies} \label{360Mbps}
\end{figure}

When the capacity of the backhaul per base-station increases to 
$270$Mbps, the proposed per-base-station SIC schemes produce a
further performance gain as compared to the $180$Mbps case.  As can be
seen in Fig.~\ref{270Mbps}, the $50$-percentile rate becomes $2.6$Mbps
for the per-base-station SIC scheme without Wyner-Ziv compression, and
$3.05$Mbps for the per-base-station SIC scheme with Wyner-Ziv
compression. However, the gap between the two curves becomes smaller.
As the capacity of the backhaul per base-station further increases to $360$Mbps,
Fig.~\ref{360Mbps} shows that the per-base-station SIC
schemes now only perform marginally better than the $270$Mbps case. 
This is when the benefit of multicell SIC starts to saturate. 

Further, as shown in Fig.~\ref{360Mbps}, the CDF curves for
the SIC schemes with and without Wyner-Ziv compression are very close
to each other for the $360$Mbps backhaul case. Thus, the
benefit of performing Wyner-Ziv compress-and-forward relaying at
the base-stations becomes negligible when the capacities of backhaul links
are high, thus confirming our earlier theoretical analysis.

In order to quantitatively evaluate the performance gain brought by
the centralized processor, Table~\ref{table_sumrate} shows the average
per-cell sum rate obtained by different schemes. The baseline
scheme gives an average per-cell sum rate of $55.5$Mbps. By utilizing
a high-capacity backhaul with Wyner-Ziv compression, up to 94\% sum
rate improvement can be obtained. Note that although the rate
improvement without Wyner-Ziv is lower than that with Wyner-Ziv,
considerable performance gains in the range of 37\% to 84\% can still be
obtained without Wyner-Ziv coding.  As noted before, the gain due to
the backhaul saturates at around $270$Mbps. 

\begin{figure}[t]
\centering
\includegraphics[width=3.4in]{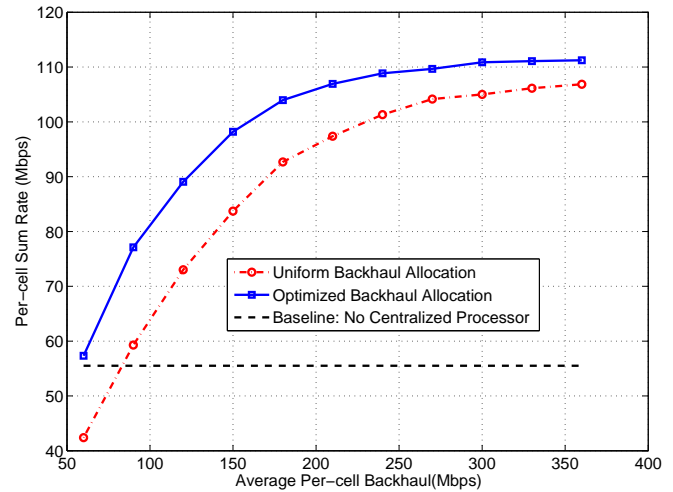}
\caption{Per-cell sum rate vs. average per-cell backhaul capacity 
	using per-base-station SIC with uniform and optimized backhaul
	capacity allocation}
\label{rate_vs_backhaul}
\end{figure}

The above simulation is performed assuming that the backhaul capacity
is the same for each base-station and 
is uniformly allocated across the frequencies. 
It is possible to further optimize the backhaul capacity allocation 
using (\ref{optimal_allocation}) of Theorem \ref{rate_allocation_theo}. 
In the next set of simulations, we choose $\alpha$ in
(\ref{optimal_allocation}) to satisfy an average backhaul constraint
across the cells, and present the resulting performance with optimized
backhaul allocation in Table~\ref{table_sumrate} and
Fig.~\ref{rate_vs_backhaul}. It can be seen that $120-150$Mbps
optimized backhaul already achieves about the same performance as that
of $180$Mbps uniform backhaul. Likewise, $180$Mbps optimized backhaul
already achieves about the same performance as that of $270$Mbps
uniform backhaul. Thus, the optimization of the backhaul is quite
beneficial.

Further, it can be seen from Fig.~\ref{rate_vs_backhaul} that under
infinite backhaul, the achieved per-cell sum rate is about $110$Mbps
for this cellular setting.  But when optimized, a finite backhaul
capacity at about 1.5 times of the user sum rate (i.e., at about
$150$Mbps) is already sufficient to achieve about $100$Mbps user sum 
rate, which is 90\% of the full benefit of uplink network MIMO. 
Note that the gain in per-cell sum rate due to the
optimization of the backhaul becomes smaller as the backhaul capacity
increases, due to the fact that increasing the backhaul
capacity eventually offers diminishing return.

\begin{table*}[t]
  \centering
  \caption{Improvement in per-cell sum rate over 19 cells, 3 sectors 
	per cell, 10 users per sector. Cell diameter is 600m. 
	Baseline per-cell sum rate is $55.5$Mbps}
  \begin{tabular}{@{} cccccc @{}}
    \toprule
    Per-cell Average & Backhaul Capacity & Per-base-station SIC & Improvement &
	Per-base-station SIC & Improvement \\
    Backhaul (Mbps)& Allocation & without WZ (Mbps) & over Baseline (\%) 
	&  with WZ (Mbps)& over Baseline (\%) \\
    \midrule
    $180$ & uniform & $75.3$ & $37\%$ & $92.7$ & $67\%$ \\
    $270$ & uniform & $93.2$ & $68\%$ & $104.2$ & $88\%$ \\
    $360$ & uniform & $102.3$ & $84\%$ & $107.6$ & $94\%$ \\
    $120$ & optimized & $75.1$ & $35\%$ & $89.1$ & $60\%$ \\
    $180$ & optimized & $83.3$ & $50\%$ & $102.9$ & $85\%$ \\
    \bottomrule
  \end{tabular}
  \label{table_sumrate}
\end{table*}

%

Finally, we mention that the performance gain presented here is
idealistic because the achievable rates are computed 
using information theoretical expressions assuming ideal coding,
modulation, and perfect CSI. In addition, all users in the 19-cell
cluster are assumed to participate in cooperative multicell
processing, and no out-of-cluster interference is accounted for. The
results in this paper nevertheless serve as upper bound to what is
achievable in an uplink network MIMO system.

\section{Conclusion}
This paper presents an information-theoretical study of a novel uplink
multicell processing scheme employing the
compress-and-forward technique with the per-base-station SIC
receiver structure for the uplink of a network MIMO system,
in which the base-stations are connected to a centralized processor
with finite capacity backhaul links. The main advantage of the
proposed schemes is that it achieves significant performance gain over
the conventional scenario where no centralized processor is deployed,
while having an achievable rate region which is easily computable, and
that it leads to an architecture that is more amendable to practical
implementations than the joint decoding scheme.  Furthermore,
theoretical analysis shows that the proposed per-base-station SIC
scheme is within a constant gap to the sum capacity for a class of
Wyner channel models.

The results of this paper also show that when employing the proposed
per-base-station SIC scheme, the capacities of the backhaul links
should scale with the logarithm of the SINR at each base-station, 
both from a point of view of approaching the theoretical maximum SIC
rate with unlimited backhaul, as well as for maximizing the overall
sum rate subject to a total backhaul rate constraint. Numerical
simulations 
reveal that significant
sum-rate gain can be obtained by the proposed SIC-based schemes
with modest backhaul capacity requirement.

%

\section*{Acknowledgements}
The authors would like to thank Dimitris Toumpakaris for helpful discussions
and valuable comments.

\bibliography{IEEEabrv,./ref/main}


\begin{IEEEbiography}{Lei Zhou}
(S'05-M'12) received the B.E. degree in Electronics Engineering
from Tsinghua University, Beijing, China,
in 2003 and M.A.Sc. and Ph.D. degrees in Electrical and Computer
Engineering from the University of Toronto, ON, Canada, in 2008 and 2012
respectively. During 2008-2009, he was with Nortel Networks, Ottawa, ON,
Canada. Since 2012, he has been with Qualcomm Technologies Inc., Santa
Clara, CA. His research interests include multiterminal information
theory, wireless communications, and signal processing.

He is a recipient of the Shahid U.H. Qureshi Memorial Scholarship in 2011,
the Alexander Graham Bell Canada Graduate Scholarship in 2011, and the
Chinese government award for outstanding self-financed students abroad in
2012.
\end{IEEEbiography}

\begin{biography}{Wei Yu}
    (S'97-M'02-SM'08) received the B.A.Sc. degree in Computer Engineering and
    Mathematics from the University of Waterloo, Waterloo, Ontario, Canada in 1997
    and M.S. and Ph.D. degrees in Electrical Engineering from Stanford University,
    Stanford, CA, in 1998 and 2002, respectively. Since 2002, he has been with the
    Electrical and Computer Engineering Department at the University of Toronto,
    Toronto, Ontario, Canada, where he is now Professor and holds a
    Canada Research Chair in Information Theory and Wireless Communications. His
    main research interests include information theory, optimization,
    wireless communications and broadband access networks.

    Prof. Wei Yu currently serves as an Associate Editor for
    {\sc IEEE Transactions on
    Information Theory}. He was an Editor for {\sc IEEE Transactions on Communications} (2009-2011), an Editor for {\sc IEEE Transactions on Wireless
    Communications} (2004-2007), and a Guest Editor for a number of
    special issues for the {\sc IEEE Journal on Selected Areas in
    Communications} and the {\sc EURASIP Journal on Applied Signal Processing}.
    He is member of the Signal Processing for Communications and Networking
    Technical Committee of the IEEE Signal Processing Society.
    He received the IEEE Signal Processing Society Best Paper Award in 2008,
    the McCharles Prize for Early Career Research Distinction in 2008,
    the Early Career Teaching Award from the Faculty of Applied Science
    and Engineering, University of Toronto in 2007, and the Early Researcher
    Award from Ontario in 2006.
\end{biography}

\end{document}